# On the origin of superlattice stacking faults nucleation via climb of Frank partial in CoNi-based superalloys

Zhida Liang[1,*], Yinan Cui[2], Li Wang[3,*], Xin Liu[2], Bin Liu[3], Yong Liu[3], Fengxian Liu[4,*]

1. Institute of Materials Research, Helmholtz-Zentrum hereon GmbH, 21502 Geesthacht, Germany
2. Applied Mechanics Lab., School of Aerospace Engineering, Tsinghua University, Beijing 100084, China
3. State Key Laboratory of Powder Metallurgy, Central South university, 410083 Changsha, China
4. Department of Mechanics of Solids, Surfaces and Systems, Faculty of Engineering Technology, University of Twente, Drienerlolaan 5, 7522NB Enschede, Netherlands

*Corresponding author: Zhida Liang (zhida.liang@outlook.com) Fengxian Liu (f.liu-3@utwente.nl); Li Wang (li.wang@csu.edu.cn)

## Abstract

High-temperature deformation in superalloys is governed by the cooperative glide-climb motion of dislocations. Superlattice stacking faults (SFs) in the γ′ phase are predominantly interpreted as nucleating via conservative Shockley partial glide. Here, we demonstrate that non-conservative climb of $a/3\langle 111 \rangle$ Frank partials constitutes a general and kinetically viable pathway for both superlattice intrinsic (SISFs) and extrinsic stacking faults (SESFs) formation in the L1$_2$ structure of CoNi-based superalloys during compression at 850 °C. High-resolution transmission electron microscopy reveals that Frank partials form at γ/γ′ interface can climb into the γ′ phase, generating SISFs via positive climb and SESFs via negative climb. Importantly, the negative climb-assisted nucleation of SESFs is experimentally confirmed for the first time, and the observed positive climb-assisted SISF configuration differs fundamentally from previously reported mechanisms. We show that these Frank partials originate from the reaction between a leading 30° Shockley partial and a 60° mixed dislocation on conjugate {111} planes, producing energetically stable configurations that promote subsequent climb. Energetic and kinetic analyses demonstrate that solute segregation induced reduction of SF energy provides a dominant contribution to Frank partial climb, enabling sustained climb and consequent SF expansion. Quantitative comparisons further indicate that, at elevated temperatures, solute drag-controlled Shockley glide can achieve mobilities comparable to vacancy diffusion-controlled Frank climb. These findings establish climb-assisted SF formation as a unified deformation mechanism in γ′ phase, and that both SISF and SESF expansion can proceed through Frank partial climb, providing new insights into dislocation-controlled high-temperature deformation in CoNi-based superalloys.

## 1. Introduction

The high-volume fraction of coherent, ordered γ′ precipitates in Ni- and Co-based superalloys plays a crucial role in impeding dislocation motion and enhancing high-temperature strength [1-3]. Dislocation-precipitate interactions are often accompanied by the formation of planar defects within the γ′ phase. Over the past few decades, extensive experimental and theoretical investigations have revealed that the dominant deformation mechanisms in superalloys are highly sensitive to temperature and stress state [4-13]. For example, at temperatures between 25 °C and 600 °C, the primary mechanism involves the glide of $a/2\langle 110\rangle$ ($a$: lattice constant) dislocations in the γ matrix channel and the cutting of γ′ precipitates by paired $a/2\langle 110\rangle$ dislocations on the same glide plane, connected by an antiphase boundary (APB) [5, 6]. At temperatures above 900 °C, dislocation climb becomes dominant, with individual $a/2\langle 110\rangle$ dislocations bypassing γ′ precipitates [11]. At intermediate temperatures (650 - 850 °C), superlattice stacking faults (SFs) emerge as the primary crystallographic defects in γ′ precipitates, significantly contributing to precipitate shearing during deformation [4, 7-10, 12, 13].

The superlattice SFs in Co-based and Ni-based superalloys are generally classified into superlattice intrinsic stacking faults (SISFs), complex intrinsic stacking faults (CISFs), and superlattice extrinsic stacking faults (SESFs). The nature of these superlattice SFs is governed by the Burgers vectors of the associated leading partial dislocations (LPDs). For instance, leading Shockley partials with $a/6\langle 112\rangle$ Burgers vector lead to unfavorable Al-Al and Ni-Ni bonding, making the formation of CISFs energetically costly, comparable to that of APBs. In contrast, leading Shockley superpartials with an $a/3\langle 112\rangle$ Burgers vector do not produce such unfavorable bonding, resulting in a significantly lower formation energy for SISFs, as illustrated in **Fig. 1**.

The energies associated with these superlattice SFs play a critical role in determining the yield strength of superalloys, particularly when deformation involves the shearing of γ′ precipitates [14-15]. A quantitative understanding of the formation mechanisms of superlattice SFs is essential for elucidating the high-temperature deformation pathways and strengthening mechanisms in superalloys, particularly in the intermediate-temperature regime where SF-mediated shearing becomes dominant.

Conventionally, the nucleation of SISFs and SESFs is attributed to the glide of Shockley partials on {111} planes. However, a fundamentally different pathway, where SFs form through the non-conservative climb motion of sessile Frank partials, has been proposed since the early work of Kear *et al.* (1970s) [16, 17]. Despite the long-standing nature of this proposal, direct experimental evidence supporting this mechanism has been still limited. Only recently has SISF nucleation via Frank partial climb been experimentally observed by Lenz *et al.* [10], whereas direct observations confirming SESF nucleation through the same mechanism are still lacking.

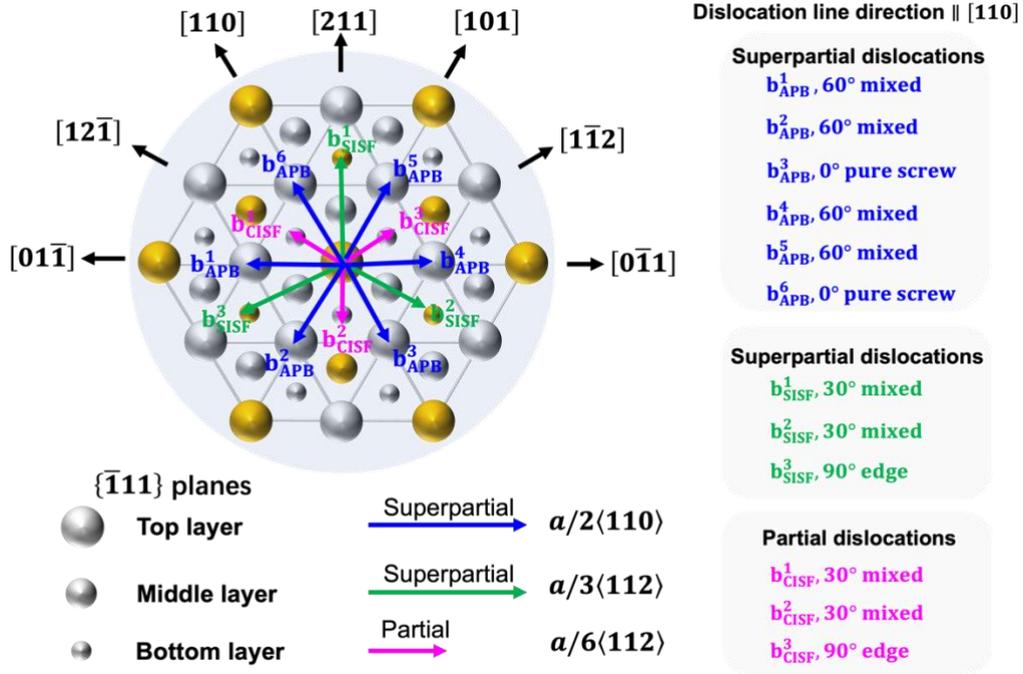

**Fig. 1. The leading dislocation Burgers vector for creating different types of stacking faults (SISFs, CISFs and APBs) in L1$_2$ structured Ni$_3$Al lattice.** Three successive $\{\bar{1}11\}$ planes are shown schematically. Large, medium and small circles represent atoms in upper, middle and lower planes correspondingly. The grey balls represent Al atoms and the yellow balls represent Ni atoms. The 6 blue arrows represent leading superpartial dislocations $a/2\langle110\rangle$ to create APBs. The 3 green arrows represent leading Shockley superpartial dislocations $a/3\langle112\rangle$ to create SISFs. The 3 magenta arrows represent leading Shockley partial dislocations $a/6\langle112\rangle$ to create CISFs.

In this work, we report two critical findings: (i) we identify a distinct positive-climb-assisted SISF nucleation pathway, differing from that proposed by Lenz *et al.* [10] both in the origin of the Frank partial and in the sequence of dislocation interactions leading to SISF nucleation, along with a systematic analysis of the dominant climb driving force that highlights the critical role of solute segregation; (ii) we provide the first direct experimental observation of SESF nucleation induced by negative Frank partial climb. These results reveal a new class of non-conservative faulting mechanisms, offering fresh insights into the dislocation dynamics governing high-temperature deformation in superalloys.

## 2. Background

As established in **Section 1**, superlattice SFs play a central role in governing γ′-shearing mechanisms in the intermediate-temperature deformation regime. Interpreting these mechanisms requires a clear and systematic understanding of the crystallographic fault configurations that can form within the L1$_2$ structure. The ordered γ′ phase permits several distinct types of superlattice SFs, including SISF, SESF, and CISF, leading to various SF shearing modes observed experimentally. In this section, we summarise the SISF and SESF configurations reported to date, providing a foundation for understanding their formation pathways and for comparison with the Frank climb-mediated SF shearing mechanisms presented in this study.

## 2.1. Superlattice intrinsic stacking faults (SISF)

To date, there are 3 different SISFs configurations which have been reported, illustrated in **Fig. 2 (a)-(c),** which are described as follows:

(1) The propagation of the mode **SISF-1**, as illustrated in **Fig. 2(a)**, was firstly linked to the formation of $a/3\langle 112\rangle$ Shockley superpartial dislocation, as firstly proposed by Kear *et al.* [18]. The reaction in **Eq. (1)** involves two distinct partial dislocations: a leading 30° Shockley superpartial $a/3[211]$ (**30°, 2Dβ**) and a trailing 30° Shockley partial $a/6[211]$ (**30°, Dβ**), as illustrated in **Fig. 2(a)**.

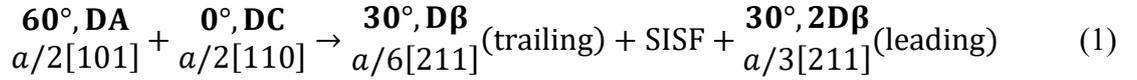

$$\underset{a/2[101]}{\overset{\mathbf{60°, DA}}{}} + \underset{a/2[110]}{\overset{\mathbf{0°, DC}}{}} \rightarrow \underset{a/6[211]}{\overset{\mathbf{30°, Dβ}}{}}\text{(trailing)} + \text{SISF} + \underset{a/3[211]}{\overset{\mathbf{30°, 2Dβ}}{}}\text{(leading)} \quad (1)$$

where the trailing Shockley partial (**30°, Dβ**) remains pinned at the γ/γ' interface.

This configuration has long been a subject of debate, and four distinct interpretations of reaction (1) have been reported in the literatures [7, 9, 19-20], as illustrated in **Fig. A.1 (Appendix 1)**. In this SISF configuration, there are no energetically unfavorable atomic bonds (*e.g.*, Ni-Ni or Al-Al), as verified by the atomic displacement analysis in **Fig. 1**. Consequently, the formation energy of an SISF associated with the leading $a/3\langle 112\rangle$ superpatial is markedly lower than that of the CISF and APB.

(2) Smith *et al.* [22] reported an SISF configuration characterized by a leading 90° $a/6\langle 112\rangle$ partial and a trailing 0° $a/6\langle 112\rangle$ partial, denoted as **SISF-2**, as illustrated in **Fig. 2(b)**. This configuration originates from the dissociation of a 60° $a/2\langle 110\rangle$ perfect dislocation. The corresponding dislocation reaction is given as follows:

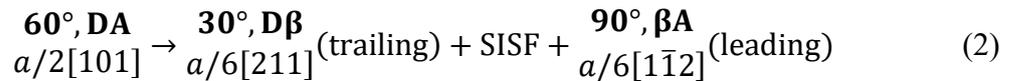

$$\underset{a/2[101]}{\overset{\mathbf{60°, DA}}{}} \rightarrow \underset{a/6[211]}{\overset{\mathbf{30°, Dβ}}{}}\text{(trailing)} + \text{SISF} + \underset{a/6[1\bar{1}2]}{\overset{\mathbf{90°, βA}}{}}\text{(leading)} \quad (2)$$

However, the existence of an $a/6\langle 112\rangle$ Shockley partial within the ordered L1$_2$ structure is energetically unfavorable, as it would produce a CISF associated with high-energy Ni-Ni or Al-Al bonding, based on description in **Fig. 1**. To alleviate this instability, local elemental segregation, particularly of Co and Cr, occurs around the CISF region, promoting atomic reordering and the subsequent formation of an SISF [8].

(3) Recent experimental observations demonstrated, for the first time that, the SISF formation can also be promoted by positive Frank partial climbing [10], as shown in **Fig. 2(c)**. Note that, a positive climb is defined as the climb motion that retracts the extra half plane of an edge dislocation. In this mechanism, two dissimilar matrix dislocations react at the γ/γ' interface to form a Lomer-type perfect dislocation (**90°, AB**), which subsequently dissociate into a 90° Frank partial (**90°, Bβ**) that climbs into the γ' phase, and a 90° Shockley partial ($90°, \boldsymbol{\beta A}$) remaining pinned at the γ/γ' interface. The corresponding two-step dislocation reaction is:

$$\underset{a/2[01\bar{1}]}{60°, AC} + \underset{a/2[\bar{1}01]}{60°, CB} \rightarrow \underset{a/2[\bar{1}10]}{90°, AB} \rightarrow \underset{a/6[\bar{1}1\bar{2}]}{90°, A\beta}(\text{trailing}) + \text{SISF} + \underset{a/3[\bar{1}11]}{90°, \beta B}(\text{leading}) \quad (3)$$

However, evaluation based on Burgers vector magnitudes indicates that this reaction needs to overcome high energy barrier due to the increase in total energy.

(4) In contrast, the SISF nucleation pathway identified in this work involves a different origin of the Frank partial, despite also proceeding through positive climb, as shown **Fig. 2(d)**. Instead of emerging from the dissociation of a Lomer-type perfect dislocation, the Frank partial forms through the sequential interaction of one 90° edge Shockley partial ($30°, \boldsymbol{\beta C}$) with intrinsic stacking fault (ISF) and one 60° mixed perfect dislocation ($60°, \boldsymbol{CB}$) from its conjugate {111} planes. The first reaction is one pure screw dislocation splitting:

$$\underset{a/2[110]}{0°, DC} \rightarrow \underset{a/6[211]}{30°, D\beta}(\text{trailing}) + \text{ISF} + \underset{a/6[12\bar{1}]}{30°, \beta C}(\text{leading}) \quad (4.a)$$

Then, one 60° mixed perfect dislocation ($60°, \boldsymbol{CB}$) approaches the leading Shockley partial ($30°, \boldsymbol{\beta C}$) at the γ/γ' interface:

$$\underset{a/6[12\bar{1}]}{30°, \beta C}(\text{leading}) + \underset{a/2[\bar{1}01]}{60°, CB} \rightarrow \underset{a/3[\bar{1}11]}{90°, \beta B}(\text{leading}) \quad (4.b)$$

The 90° Frank partial ($90°, \boldsymbol{B\beta}$) climbs towards the γ' phase via vacancy absorption/emission, while the 30° Shockley partial ($30°, \boldsymbol{D\beta}$) remain pinned at the γ/γ' interface. The overall dislocation reaction is:

$$\underset{a/2[110]}{0°, DC} + \underset{a/2[\bar{1}01]}{60°, CB} \rightarrow \underset{a/6[211]}{30°, D\beta}(\text{trailing}) + \text{SISF} + \underset{a/3[\bar{1}11]}{90°, \beta B}(\text{leading}) \quad (5)$$

This newly proposed mechanism differs from the previously reported pathway both in the origin of the Frank partial and in the sequence of dislocation interactions leading to SISF nucleation. Experimental observations and energetic analysis supporting this configuration are presented in the following sections.

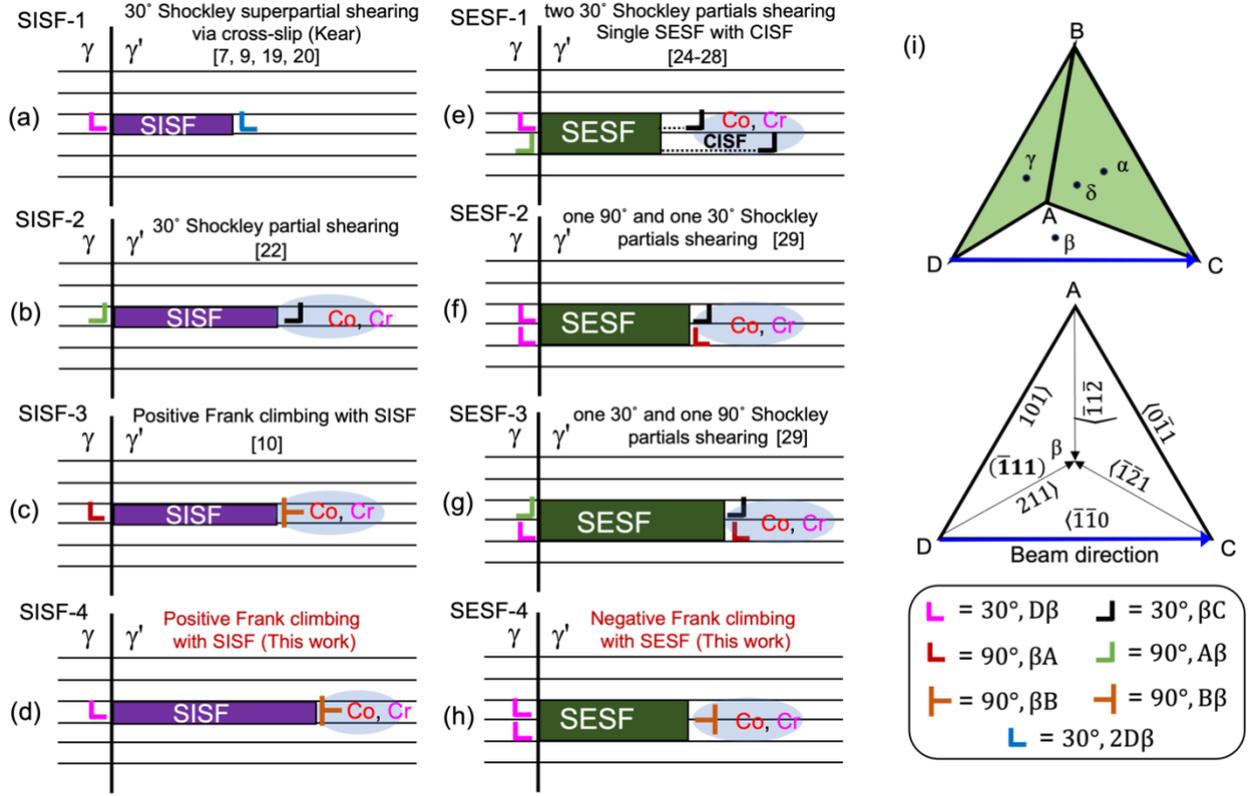

**Fig. 2. Summary of potential superlattice SFs shearing modes in superalloys.** (a)-(d) The four different SISF configurations in superalloys, including the SISF nucleation via positive Frank partial climb. (e)-(h) The four different SESF configurations in superalloys. (i) 3D and 2D Thompson tetrahedron. **DC** = [110] is dislocation line direction, parallel to the electron beam direction in TEM. Plane **ADC** is ($\bar{1}$11).

## 2.2. Superlattice extrinsic stacking faults (SESF)

The earliest mechanism for SESF formation was proposed by Kear *et al.* [23], who hypothesized that the concerted glide of three Shockley partial dislocations with different signs can produce the $D0_{24}$ local ordering characteristic of a two-layer SESF within a $L1_2$ precipitate. An alternative view, following Kolbe model [13], is that two-layer CISFs first form at the γ/γ′ interface and subsequently transform into an SESF through local elemental segregation and atomic reordering. This pathway has since received experimental support [5, 24-28]. On this basis, three glide-mediated SESF modes have been reported in the literatures to date [24-29].

(1) The mode **SESF-1** has been studied in a number of publications [24-28]. The reaction involves dissociation of one pure screw 0° dislocation and one 60° mixed dislocations, producing two CISFs on adjacent {111} plane, as illustrated in **Fig. 2(e)**:

$$\underset{a/2\,[01\bar{1}]}{\mathbf{60°, AC}} \rightarrow \underset{a/6[\bar{1}1\bar{2}]}{\mathbf{90°, A\beta}}(\text{trailing}) + \mathbf{CISF} + \underset{a/6[12\bar{1}]}{\mathbf{30°, \beta C}}(\text{leading}) \qquad (6.\text{a})$$

$$\begin{array}{c} 0°, \mathbf{DC} \\ a/2\,[110] \end{array} \rightarrow \begin{array}{c} 30°, \mathbf{D\beta} \\ a/6[211] \end{array}(\text{trailing}) + \mathbf{CISF} + \begin{array}{c} 30°, \mathbf{\beta C} \\ a/6[12\bar{1}] \end{array}(\text{leading}) \qquad (6.\text{b})$$

These two CISFs subsequently transform into two layers SESF via atomic reordering.

(2) Recently, **SESF-2** and **SESF-3** were identified in our previous publication [29]. In mode **SESF-2**, the LPDs consist of one 90° Shockley edge partial and one 30° Shockley partial. The trailing part is composed of two 30° Shockley partial remain pinned at the γ/γ′ interface, as illustrated in **Fig. 2(f)**.

$$\begin{array}{c} 0°, \mathbf{DC} \\ a/2\,[110] \end{array} \rightarrow \begin{array}{c} 30°, \mathbf{D\beta} \\ a/6[211] \end{array}(\text{trailing}) + \mathbf{CISF} + \begin{array}{c} 30°, \mathbf{\beta C} \\ a/6[12\bar{1}] \end{array}(\text{leading}) \qquad (7.\text{a})$$

$$\begin{array}{c} 60°, \mathbf{DA} \\ a/2\,[101] \end{array} \rightarrow \begin{array}{c} 30°, \mathbf{D\beta} \\ a/6[211] \end{array}(\text{trailing}) + \mathbf{CISF} + \begin{array}{c} 90°, \mathbf{\beta A} \\ a/6[1\bar{1}2] \end{array}(\text{leading}) \qquad (7.\text{b})$$

The angle of Shockley partials **30°, βC** and **90°, βA** is 120° and therefore these two LPDs will attract each other to form a **30°, Dβ**. Therefore, the reaction **Eq. (7)** can be written as:

$$\begin{array}{c} 0°, \mathbf{DC} \\ a/2\,[110] \end{array} + \begin{array}{c} 60°, \mathbf{DA} \\ a/2\,[101] \end{array} \rightarrow \begin{array}{c} 30°, \mathbf{2D\beta} \\ a/3[211] \end{array}(\text{trailing}) + \mathbf{SESF} + \begin{array}{c} 30°, \mathbf{D\beta} \\ a/6[211] \end{array}(\text{leading}) \qquad (8)$$

The reaction in **Eq. (8)** was ever reported in literature by Knowles and Chen [30].

(3) Another configuration is **SESF-3** [29], the LPDs comprise a 90° Shockley edge partial and a 30° Shockley partial. The trailing dislocations, consisting of another 90° Shockley edge partial and 30° Shockley partial, remain pinned at the γ/γ′ interface, as illustrated in **Fig. 2(g)**:

$$\begin{array}{c} 60°, \mathbf{AC} \\ a/2\,[01\bar{1}] \end{array} \rightarrow \begin{array}{c} 90°, \mathbf{A\beta} \\ a/6[\bar{1}1\bar{2}] \end{array}(\text{trailing}) + \mathbf{CISF} + \begin{array}{c} 30°, \mathbf{\beta C} \\ a/6[12\bar{1}] \end{array}(\text{leading}) \qquad (9.\text{a})$$

$$\begin{array}{c} 60°, \mathbf{DA} \\ a/2\,[101] \end{array} \rightarrow \begin{array}{c} 30°, \mathbf{D\beta} \\ a/6[211] \end{array}(\text{trailing}) + \mathbf{CISF} + \begin{array}{c} 90°, \mathbf{\beta A} \\ a/6[1\bar{1}2] \end{array}(\text{leading}) \qquad (9.\text{b})$$

where the reaction is two dissimilar 60° mixed dislocations splitting with two CISFs formation on neighboring {111} plane.

In all reported modes, **SESF-1**, **SESF-2** and **SESF-3**, the nucleation of SESFs is mainly driving by Shockley partial glide, the formation process is therefore glide-controlled and conservative.

(4) In contrast, in the present work, we identify a new SESF nucleation mechanism, designated as **SESF-4**, which differs fundamentally from previously reported LPD-type and its SESF expansion mode. As shown in **Fig. 2(h)**, **SESF-4** forms from the negative climb of a Frank partial dislocation (LPD) in CoNi-based superalloys. Distinct from the glide-controlled pathways, this represents the first direct experimental observation of SESF formation within the $L1_2$ structure via a non-conservative climb process. In the following sections, we further elucidate how the Frank partial forms at the γ/γ′ interface and subsequently climbs into the γ′ phase to create the **SESF-4** configuration.

## 3. Materials and methods

The as-cast alloy was homogenized at 1250 °C for 24 hours, followed by aging at 900 °C in air for 220 hours, and subsequently air-cooled to room temperature. The nominal composition of the superalloy was Co-35Ni-15Cr-5Al-5Ti-2Mo-1W-0.1B (at. %). The lattice parameters of the γ′ and γ phases were measured to be $a_{\gamma'}$ = 0.3580 nm and $a_{\gamma}$ = 0.3565 nm, respectively. Compression tests were conducted using a strain-controlled, closed-loop MTS 810 testing machine (MTS Systems Corporation). Compression specimens, with a gauge length of 7.5 mm and a diameter of 5 mm, were machined from the standard heat-treated material. The tests were performed at a strain rate of $10^{-4}$ s$^{-1}$ under 850 °C. Deformation during testing was monitored using an extensometer with a 21 mm gauge length, which provided feedback for strain control.

For transmission electron microscopy (TEM) sample preparation, 3 mm disks were sectioned from the deformed samples and mechanically polished to a thickness of approximately 75 μm. These disks were subsequently thinned to electron transparency using a twin-jet electropolishing unit with Struers A3 electrolyte, operated at a voltage of 32 V and a temperature of -38 °C. The traditional TEM in dark field mode was done in Talos 200i. High-resolution high-angle annular dark field (HAADF) imaging and energy-dispersive X-ray spectroscopy (EDS) were carried out using a Thermo Fisher Scientific Themis Z and Spectra 300 transmission electron microscope, operated at 300 kV and equipped with a probe corrector. These analyses provided atomic-resolution insights into the defect structures in the material. High-resolution STEM (HRSTEM) geometrical phase analysis (GPA) strain analysis was realized by a plugin [31] to the commercial software DigitalMicrograph (DM3). In addition, HRSTEM images were analysed by center of symmetry (COS) method with MATLAB codes [32].

Additional finite element (FE) calculations were performed to evaluate the magnitude of stresses arising from lattice misfit, applied loading, and interfacial dislocation pile-ups. These calculations are intended to illustrate the possible magnitude and sign of local stresses acting on Frank partials and are described in detail in the **Supplementary Material 1, Fig. S4**.

## 4. Results

In this section, we first identify the coexistence of SISFs and SESFs within the γ′ precipitates, highlighting their distinct structural characteristics revealed by HRSTEM analysis. Subsequently, we investigate the nucleation mechanisms of these faults, demonstrating that SISFs originate through the positive climb of Frank partial dislocations, whereas SESFs form via the negative climb of Frank partials. Furthermore, we show the observation of ISFs at the γ/γ′ interface and in the γ matrix channel, which act as precursors for subsequent SISF and SESF formation through local dislocation reactions and climb-assisted processes. Finally, elemental segregation is quantitatively analyzed along both the fault planes (SISFs and SESFs) and their associated leading dislocations (Frank and Shockley partials), enabling a direct comparison of their compositional differences in Cottrell atmosphere.

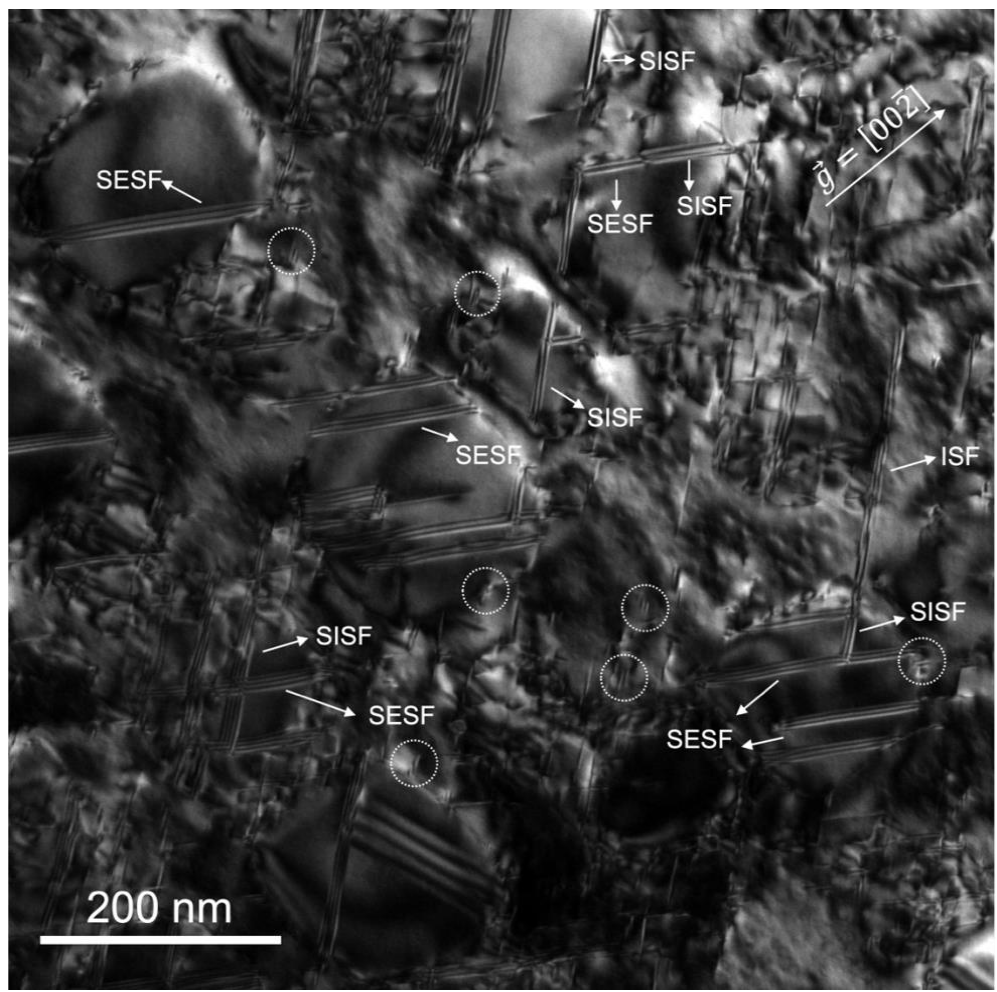

**Fig. 3.** Overview DF-TEM image of the deformed sample at 3% strain in [110] beam direction (SFs edge-on view direction), showing high densities of SISFs and SESFs in γ′ phases. Occasionally, SFs were observed in both the γ matrix channel and at the γ/γ′ interface (marked by white circle).

**Fig. 3** presents DF-TEM overview of the deformed microstructure in a specimen with a [110] surface normal, obtained after compressive deformation to 3% plastic strain. Numerous planar defects, primarily SFs, are observed to terminate within the γ′ precipitates. According to Edington' description [33, 34], for DF imaging using a {200} reflection, if the g-vector is oriented toward the bright outermost fringe, the SF is intrinsic; conversely, if the g-vector is oriented away, the fault is extrinsic. Based on this criterion, both SISFs and SESFs are identified to coexist within the same grain, revealing the complexity of fault nucleation and propagation under compressive loading. In addition, some SFs were occasionally observed in the γ matrix channel and at the γ/γ′ interface. The intrinsic type of these SFs is determined by HRSTEM in **Section 4.3**.

**4.1. The SISF configuration with leading Frank partial in the γ′ phase**

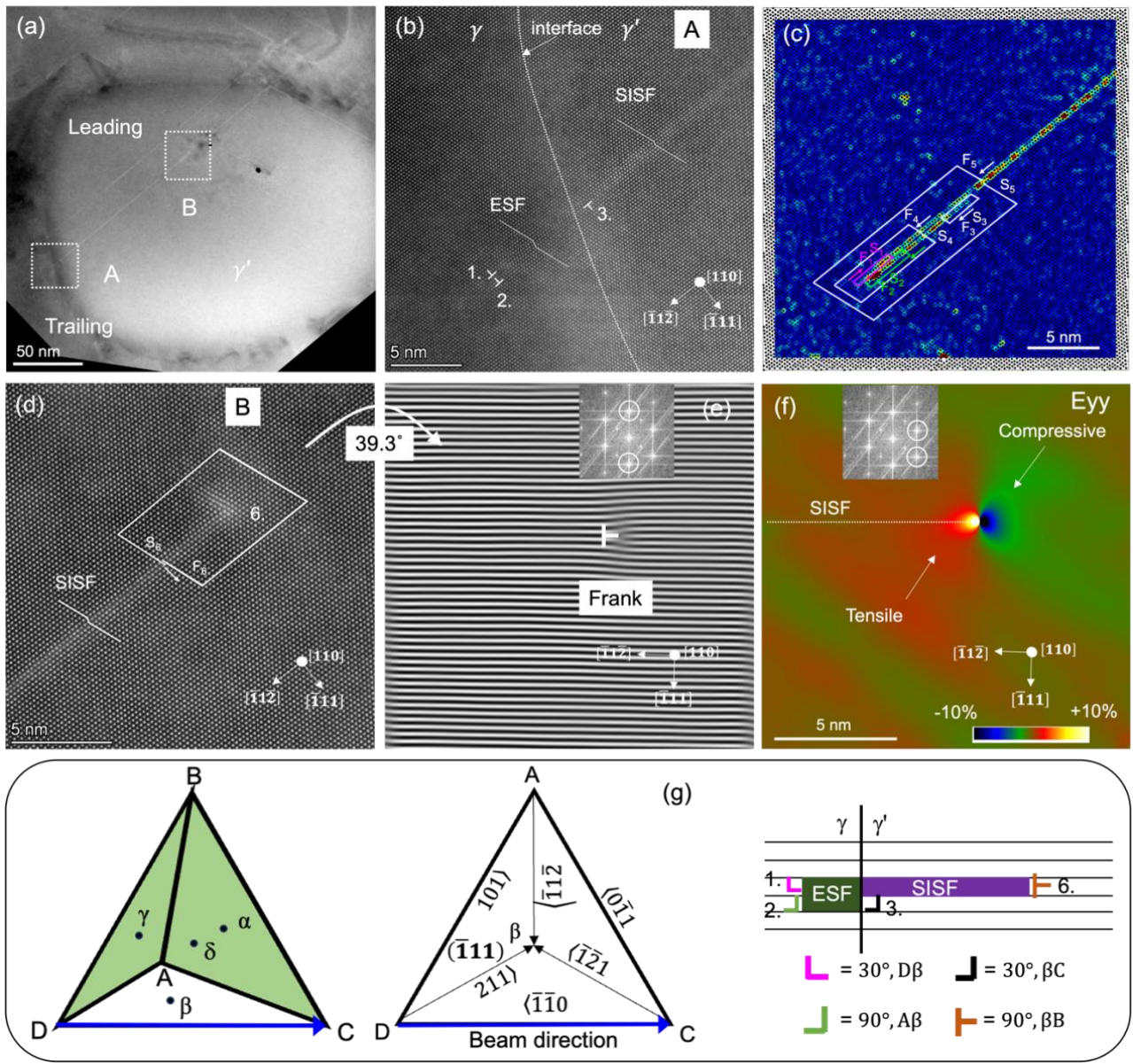

**Fig. 4. The nucleation of SISF in the γ′ phase via Frank positive climb**. (a) Low magnification HAADF-STEM micrograph of several SFs terminating inside a γ′ precipitate. (b) Atomic resolution HAADF-STEM image of trailing part of the SESF inside a γ′ precipitate and the Shockley partial dislocation remain pinned at the γ/γ′ interface. (c) The COS-visualization of image **Fig. 4(b)**. (d) Atomic resolution HAADF-STEM image of the leading part of an SESF terminating inside a γ′ precipitate and the leading Frank partial dislocation. (e) Bragg filtered image of the SESF using two reflections $g = \pm (\bar{1}11)$ and the leading Frank partial showing an additional half-plane absence. (f) GPA $E_{yy}$ strain analysis maps of the strain component parallel to the {111} crystal planes using two reflections $g = (1\bar{1}1)$ and $g = (002)$. (g) Schematic illustration of Frank-type SISF substructure with a leading Frank partial and three trailing Shockley partials based on experimental analysis.

**Fig. 4** presents direct evidence of the SISF nucleation within the γ′ phase via the positive climb of a Frank partial. The HAADF-STEM image in **Fig. 4(a)** shows several SFs and their terminations, where the leading (region B) and trailing (region A) parts of the SF of interest are highlighted for detailed structural analysis. The stacking sequence of the $(\bar{1}11)$ planes in the high-resolution HAADF-STEM images (**Fig. 4(b)** and **(d)**) confirms that the fault is of the intrinsic type, corresponding to an SISF.

To further elucidate the dislocation configuration associated with the SISF, a center-of-symmetry (COS) analysis was performed. The resulting COS color map (**Fig. 4(c)**) clearly reveals the local lattice distortion and provides additional confirmation that the observed fault is indeed an SISF. In addition, an extrinsic stacking fault (ESF), located at the γ/γ′ interface, was observed in trailing part of the SISF. The leading segment contains one leading shearing dislocation, while the trailing segment comprises three trailing dislocations, two of which are pinned at the γ/γ′ interface.

Clockwise Burgers circuits were constructed around the leading and trailing dislocations to determine their Burgers vectors. The projected Burgers vectors around the trailing part are determined to be $\mathbf{b}_{1,p} = \mathbf{F}_1\mathbf{S}_1 = a/12[1\bar{1}2]$, $\mathbf{b}_{2,p} = \mathbf{F}_2\mathbf{S}_2 = a/6[\bar{1}1\bar{2}]$, $\mathbf{b}_{3,p} = \mathbf{F}_3\mathbf{S}_3 = a/12[\bar{1}1\bar{2}]$, $\mathbf{b}_{4,p} = \mathbf{F}_4\mathbf{S}_4 = a/12[\bar{1}1\bar{2}]$ and $\mathbf{b}_{5,p} = \mathbf{F}_5\mathbf{S}_5 = a/6[\bar{1}1\bar{2}]$, where $\mathbf{b}_{3,p} - \mathbf{b}_{4,p} = \mathbf{b}_{5,p}$. The actual Burgers vectors of the trailing dislocations (1, 2, and 3) are therefore $\mathbf{b}_1 = a/6[211]$, $\mathbf{b}_2 = a/6[\bar{1}1\bar{2}]$ and $\mathbf{b}_3 = a/6[12\bar{1}]$, corresponding respectively to (**30°, Dβ (1)**), (**90°, Aβ (2)**) and (**30°, βC (3)**) dislocations in the Thompson tetrahedron.

For the leading partial, bounding the SISF on the right side (**Fig. 4(c)**), the projected Burgers vector is $\mathbf{b}_{6,p} = \mathbf{F}_6\mathbf{S}_6 = a/3[\bar{1}11]$. Since the Frank dislocation is a 90° edge dislocation, the actual Burgers vector is $\mathbf{b}_6 = \mathbf{b}_{6,p} = a/3[\bar{1}11]$, corresponding to a (**90°, βB**) dislocations in the Thompson tetrahedron. The dislocation line direction, $\mathbf{l} = [110]$, is perpendicular to the Burgers vector, confirming that the Frank partial undergoes non-conservative climb motion characteristic of a pure edge dislocation. Further explanations of distinguishing between 30° mixed-type and 90° edge-type partial dislocations, as well as the method for determining the real Burgers vectors from the projected ones in HRSTEM images, are provided in **Supplementary Materials 1**, **Fig. S1**.

A Bragg-filtered image (**Fig. 4(e)**) shows the absence of an extra half-plane near the fault. Combined with the presence of an SISF, this observation confirms that the leading dislocation is a Frank partial. Geometric phase analysis (GPA), shown in **Fig. 4(f)**, reveals long-range strain fields emanating from the Frank partial, with a tensile strain region adjacent to the SISF where the extra half-plane is missing. Based on these observations, the schematic in **Fig. 4(g)** illustrates the proposed substructure of the Frank-type SISF. This configuration demonstrates that SISFs can nucleate through the positive climb of Frank dislocations, linking point-defect diffusion-controlled climb processes to planar fault evolution within the γ′ phase.

## 4.2. The SESF configuration with leading Frank partial in the γ′ phase

**Fig. 5** presents direct evidence of SESF nucleation within the γ′ phase through the negative climb of a Frank dislocation. The HAADF-STEM micrograph in **Fig. 5(a)** displays several SFs and their terminations, with the leading (region B) and trailing (region A) segments of the selected SESF highlighted for detailed analysis. The leading segment contains one shearing dislocation, whereas the trailing segment consists of two dislocations pinned at the γ/γ′ interface. High-resolution HAADF-STEM images (**Fig. 5(b, c)**) reveal the stacking sequence of the $(\bar{1}11)$ planes, confirming the extrinsic nature of the fault. Based on the Burgers circuit and the finish-start (FS) right-hand convention, the projected Burgers vector for the trailing part is $\mathbf{b_{1,p}} = \mathbf{F_1S_1} = a/6[1\bar{1}2]$, corresponding to an actual vector $\mathbf{b_1} = 2\mathbf{D\beta} = a/3[211]$, which consists of two 30° Shockley partials. For the leading segment (**Fig. 5(c)**), the projected Burgers vector is $\mathbf{b_{2,p}} = \mathbf{F_2S_2} = a/3[1\bar{1}\bar{1}]$. Because Frank partial is 90° edge dislocation, the actual Burgers vector, deduced from the projection, is $\mathbf{b_2} = \mathbf{b_{2,p}} = a/3[1\bar{1}\bar{1}]$ corresponding to one **(90°, Bβ)** edge-type dislocation in the Thompson tetrahedron.

The dislocation line $l = [110]$ is perpendicular to the Burgers vector, indicating non-conservative negative climb motion of a pure edge dislocation. The center-of-symmetry (COS) map (**Fig. 5(d)**) further confirms the extrinsic character of the fault, while the Bragg-filtered image (**Fig. 5(e)**) shows an inserted half-plane near the SESF, consistent with a Frank partial. Geometric phase analysis (GPA) (**Fig. 5(f)**) reveals long-range strain fields emanating from the Frank dislocation and compressive strain adjacent to the inserted half-plane, opposite to that observed for SISFs. The schematic in **Fig. 5(g)** summarizes the Frank-type SESF substructure, where the LPD is a **(90°, Bβ)** Frank partial and the trailing dislocations are two **(30°, Dβ)** Shockley partials. This configuration provides direct experimental evidence that SESFs can nucleate through the negative climb of Frank partial.

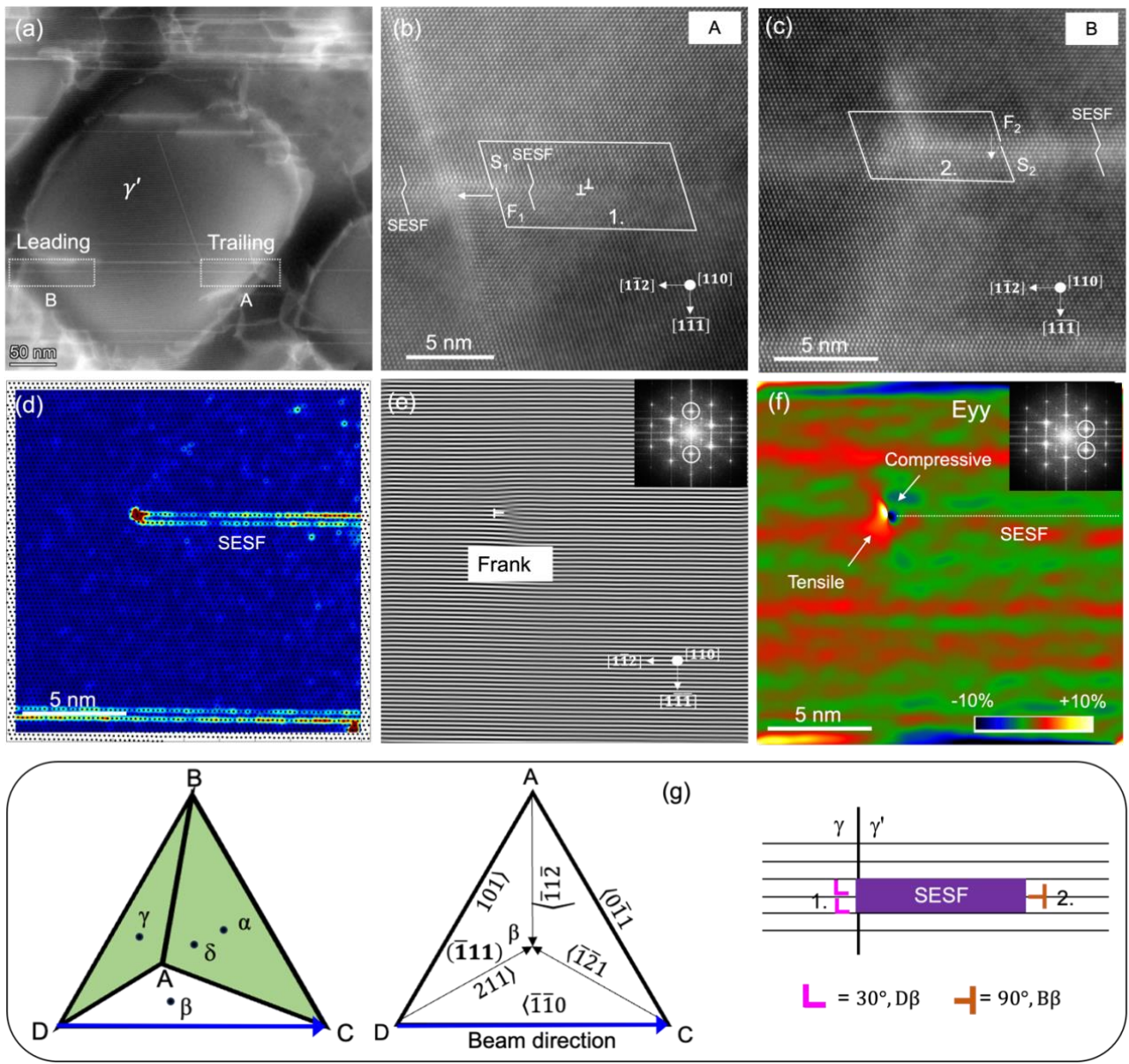

**Fig. 5. The nucleation of SESF in the γ′ phase via Frank negative climb**. (a) Low magnification HAADF-STEM micrograph of several SFs terminating inside a γ′ precipitate. (b) Atomic resolution HAADF-STEM image of trailing part of the SESF inside a γ′ precipitate and the Shockley partial dislocation remain pinned at the γ/γ′ interface. (c) Atomic resolution HAADF-STEM image of the leading part of an SESF terminating inside a γ′ precipitate and the leading Frank partial dislocation. (d) The COS-visualization of image **Fig. 5(c)**. (e) Bragg filtered image of the SESF using two reflections $\boldsymbol{g} = \pm(\bar{1}11)$ and the leading Frank partial showing an additional half-plane inserted. (f) GPA $E_{yy}$ strain analysis maps of the strain component parallel to the {111} crystal planes using two reflections $\boldsymbol{g} = (\bar{1}1\bar{1})$ and $\boldsymbol{g} = (00\bar{2})$. (g) Schematic illustration of Frank-type SISF substructure with a leading Frank partial and two trailing Shockley partials based on experimental analysis.

## 4.3. The origin of ISFs at the γ/γ′ interface and in the γ matrix channel

The γ matrix channel are usually saturated with numerous dislocations, which serve as the "raw material" for the formation of $a/6\langle112\rangle$ and $a/2\langle110\rangle$ dislocations being capable of shearing γ′ the precipitates [9, 13, 35]. Kolbe [13] investigated, by *in situ* TEM at 780°C, the onset of a new deformation mechanism in Ni-based superalloys. Above 780°C, deformation proceeds through the viscous glide of paired γ-matrix dislocations with Burgers vectors of type $a/6\langle112\rangle$, which move by dragging along intrinsic stacking faults (ISFs). Bürger *et al.* [36] further confirmed, using conventional TEM, that these $a/6\langle112\rangle$ type Shockley partials originate from the splitting of $a/2\langle110\rangle$ perfect dislocations. Vorontsov *et al.* [9] proposed that the nucleation of SESFs and SISFs arises from synthetic reactions between distinct CISFs at the γ/γ′ interface. Through subsequent solute segregation and atomic reordering, two adjacent CISFs can transform into either a two-layer SESF [13, 24-28] or a single-layer SISF [9] (**story 3** of **Fig. A. 1** in **Appendix 1**). However, the character and nature of ISFs that actually form at the γ/γ′ interface remain unclear. In this section, three distinct ISFs were identified, two located at the γ/γ′ interface and one within the γ matrix channel, together with an SESF terminating inside the γ′ phase, as shown in **Figs. 6** and **7**.

### 4.3.1 The ISFs at the γ/γ′ interface

The HAADF-STEM image in **Fig. 6(a)** shows several SFs located at the γ/γ′ phase interface and one superlattice SF terminating inside the γ′ phase. The leading (region B) and trailing (region A) parts of the SFs are highlighted for detailed structural analysis. Three SFs are identified: two ISFs, labeled **SF1_ISF** and **SF2_ISF**, and one SESF, labeled **SF3_SESF**, as illustrated in **Fig. 6(b)**.

For **SF1_ISF**, shown in **Fig. 6(b),** clockwise Burgers circuits were constructed around the leading and trailing dislocations to determine their Burgers vectors. The projected Burgers vectors for the trailing part are $\mathbf{b_{1,p}} = \mathbf{F_1S_1} = a/6[\bar{1}1\bar{2}]$ and $\mathbf{b_{2,p}} = \mathbf{F_2S_2} = a/12[\bar{1}1\bar{2}]$, giving $\mathbf{b_{1,p}} + \mathbf{b_{2,p}} = a/4[\bar{1}1\bar{2}]$. The corresponding true Burgers vectors are $\mathbf{b_1} = a/6[\bar{1}1\bar{2}]$ and $\mathbf{b_2} = a/6[12\bar{1}]$, which correspond to **(90°, Aβ)** and **(30°, βC)** Shockley partials, respectively, in the Thompson tetrahedron. The **SF1_ISF** originates from the splitting of a 60° mixed perfect dislocation **AC** on the $(\bar{1}11)$ plane, denoted as **ADC**. The resulting **SF1_ISF** substructure, shown in **Fig. 6(e)**, consists of a leading **(90°, Bβ)** and a trailing **(30°, Dβ)** Shockley partials.

For **SF2_ISF**, shown in **Fig. 6(c),** clockwise Burgers circuits were also used to identify the dislocation characteristics. The projected Burgers vectors for the trailing part are $\mathbf{b_{3,p}} = \mathbf{F_3S_3} = a/12[1\bar{1}2]$ and $\mathbf{b_{4,p}} = \mathbf{F_4S_4} = a/12[\bar{1}1\bar{2}]$, giving $\mathbf{b_{3,p}} + \mathbf{b_{4,p}} = 0$. The actual Burgers vectors are therefore $\mathbf{b_3} = a/6[211]$ and $\mathbf{b_4} = a/6[12\bar{1}]$, corresponding to **(30°, Dβ)** and **(30°, βC)** Shockley partials, respectively. This ISF originates from the dissociation of a 0° pure screw perfect dislocation

**DC** on the $(\bar{1}11)$ plane, denoted as **ADC**. The **SF2_ISF** substructure, illustrated in **Fig. 6(f)**, consists of a leading **(30°, Dβ)** and a trailing **(30°, βC)** Shockley partials.

For **SF3_SESF**, the trailing and leading regions are shown in **Fig. 6(b)** and **Fig. 6(d)**, respectively. Clockwise Burgers circuits were constructed to determine the associated Burgers vectors. The projected Burgers vector for the trailing part is $\mathbf{b}_{5,p} = F_5S_5 = a/6[1\bar{1}2]$ and the corresponding true Burgers vector is $\mathbf{2D\beta} = a/3[211]$, indicating the presence of two 30° Shockley partials.

The leading part contains three Burgers circuits with projected vectors $\mathbf{b}_{6,p} = F_6S_6 = a/12[1\bar{1}2]$, $\mathbf{b}_{7,p} = F_7S_7 = a/12[\bar{1}1\bar{2}]$ and $\mathbf{b}_{8,p} = F_8S_8 = a/6[2\bar{1}1]$. The actual Burgers vectors are $\mathbf{b}_6 = a/6[211]$ (denoted **Dβ**), $\mathbf{b}_7 = a/6[12\bar{1}]$ (**βC**) and $\mathbf{b}_8 = a/6[1\bar{1}2]$ (**βA**). The vector relation $\mathbf{b}_6 = \mathbf{b}_7 + \mathbf{b}_8$ confirms that the leading configuration comprises a **(90°, βA)** and a **(30°, βC)** Shockley partial. Consequently, the **SF3_SESF** substructure (**Fig. 6(g)**) consists of two LPDs (**30°, βC** and **90°, βA**) and two trailing dislocations (**30°, 2Dβ**). This configuration supports the mechanism whereby an SESF forms via the reaction between a 0° pure screw dislocation (**DB**) and a 60° mixed dislocation (**AB**) on the $(\bar{1}11)$ plane, denoted as **ACD**.

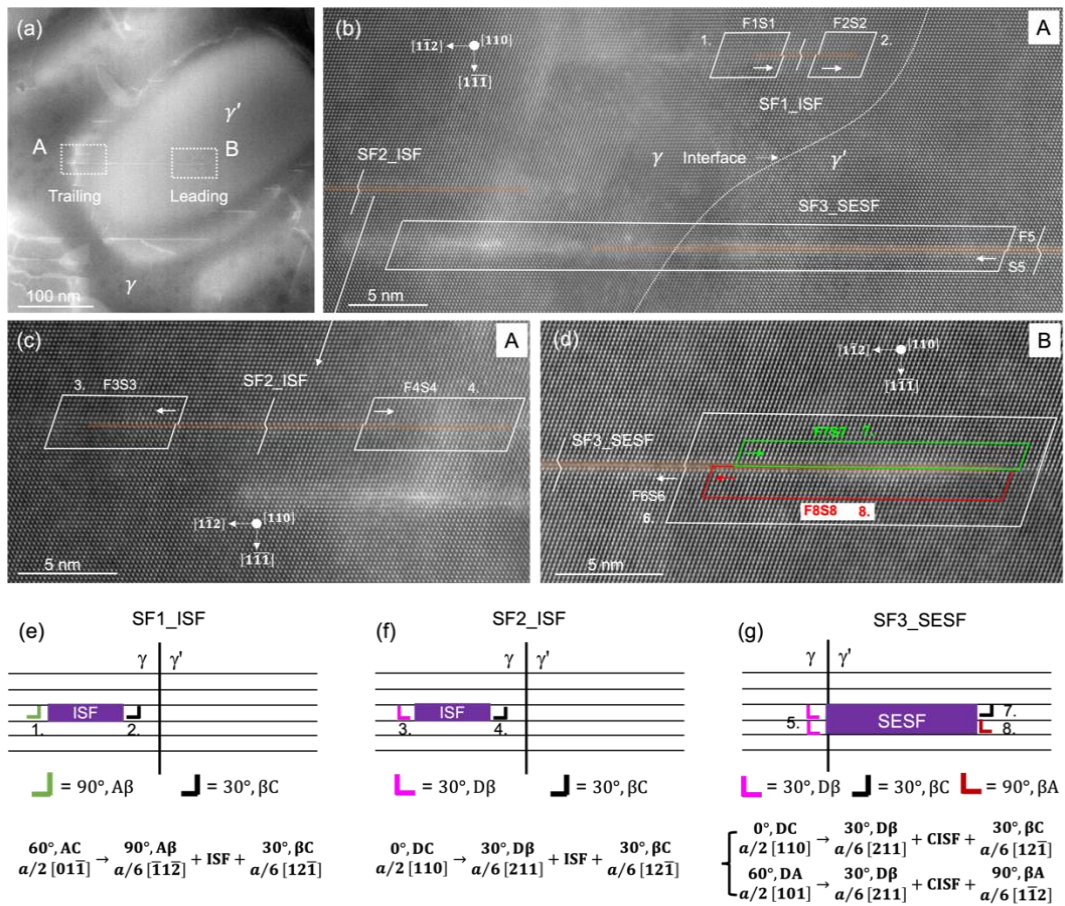

**Fig. 6. Nucleation of ISFs at the γ/γ′ interface.** (a) Several SFs located at the γ/γ′ phase interface, including one superlattice SF terminating within the γ′ phase. (b)-(c) Two ISFs observed at the γ/γ′

interface, labeled **SF1_ISF** and **SF2_ISF**, and one trailing segment of a SESF, labeled **SF3_SESF**. The two Shockley partials associated with **SF3_SESF** remain pinned at the γ/γ′ interface. (d) Atomic-resolution HAADF-STEM image showing the leading part of an SESF terminating inside a γ′ precipitate. (e) Schematic illustration of the **SF1_ISF** substructure with a leading Shockley partial (**30°, βC**) and a trailing Shockley partial (**90°, Aβ**). (f) Schematic of the **SF2_ISF** substructure showing a leading Shockley partial (**30°, βC**) and one trailing Shockley partial (**30°, Dβ**). (g) Schematic of **SF3_SESF** substructure containing two leading Shockley partials (**30°, βC** and **90°, βA**) and two trailing Shockley partials (**30°, Dβ**), as deduced from experimental analysis.

**4.3.2 The ISFs in the γ matrix channel**

The ISF defects at the γ/γ′ interface were analyzed above. In addition to those interfacial faults, SFs were also observed within the γ matrix channel, as shown in **Fig. 7(a)-(b)**. The HAADF-STEM images reveal two ISFs, labeled **SF4_ISF** and **SF5_ISF**. The leading (region B) and trailing (region A) parts of **SF4_ISF**, as well as the overall configuration of **SF5_ISF** (region C), are highlighted for detailed structural analysis.

For **SF4_ISF**, clockwise Burgers circuits were drawn around the trailing (**Fig. 7(c)**) and leading (**Fig. 7(d)**) dislocations to determine their Burgers vectors. The projected Burgers vectors for the trailing part are $\mathbf{b}_{9,p} = F_9S_9 = a/6[\bar{1}1\bar{2}]$ and $\mathbf{b}_{10,p} = F_{10}S_{10} = a/12[\bar{1}1\bar{2}]$, giving $\mathbf{b}_{3,p} + \mathbf{b}_{4,p} = a/4[\bar{1}1\bar{2}]$. The corresponding true Burgers vectors are $\mathbf{b}_9 = a/6[\bar{1}1\bar{2}]$ and $\mathbf{b}_{10} = a/6[12\bar{1}]$, which correspond to **90°, Aβ** and **30°, βC** Shockley partials, respectively. **SF4_ISF** originates from the splitting of a 60° mixed perfect dislocation **AC** on the ($\bar{1}11$) plane, denoted as **ADC**. The resulting substructure, shown in **Fig. 7(f)**, is identical to that of **SF1_ISF**.

For **SF5_ISF**, shown in **Fig. 7(d)**, clockwise Burgers circuits were similarly constructed around the leading and trailing dislocations. The projected Burgers vectors for the trailing part are $\mathbf{b}_{11,p} = F_{11}S_{11} = a/12[1\bar{1}2]$ and $\mathbf{b}_{12,p} = F_{12}S_{12} = a/6[1\bar{1}2]$, yielding $\mathbf{b}_{11,p} + \mathbf{b}_{12,p} = a/4[1\bar{1}2]$. The corresponding true Burgers vectors are $\mathbf{b}_{11} = a/6[211]$ and $\mathbf{b}_{12} = a/6[12\bar{1}]$, corresponding to **30°, Dβ** and **90°, βA** Shockley partials, respectively. **SF5_ISF** originates from the splitting of a 60° mixed perfect dislocation **AC** on the ($\bar{1}11$) plane, denoted as **ADC**. The resulting substructure, shown in **Fig. 7(g)**, consists of a leading **30°, Dβ** and one trailing **90°, βA** Shockley partials.

Based on the analyses in **Figs. 6** and **7**, three distinct ISFs were identified at the primary γ/γ′ interface and within the γ matrix channel. These ISFs originate from the dissociation of three different perfect dislocations (**60°, DA**, **60°, AC** and **0°, DC**) on the ($\bar{1}11$) plane, as represented in the Thompson tetrahedron in **Fig. 2(i)**. These ISFs serve as the "raw materials" for SESF formation. According to the model proposed by Kolbe *et al.* [13], followed by Smith *et al.* [24], Kovarik *et al.* [26-27], Karpstein *et al.* [28] and Barba *et al.* [25], two ISFs on adjacent {111} planes can shear the γ′ precipitate, forming a two-layer CISF. Through subsequent elemental diffusion and atomic reordering, the two-layer CISF transforms into a two-layer SESF. **SF3_SESF** provides a representative example of this process, originating from two ISFs interaction of **SF2_ISF** and **SF4_ISF**.

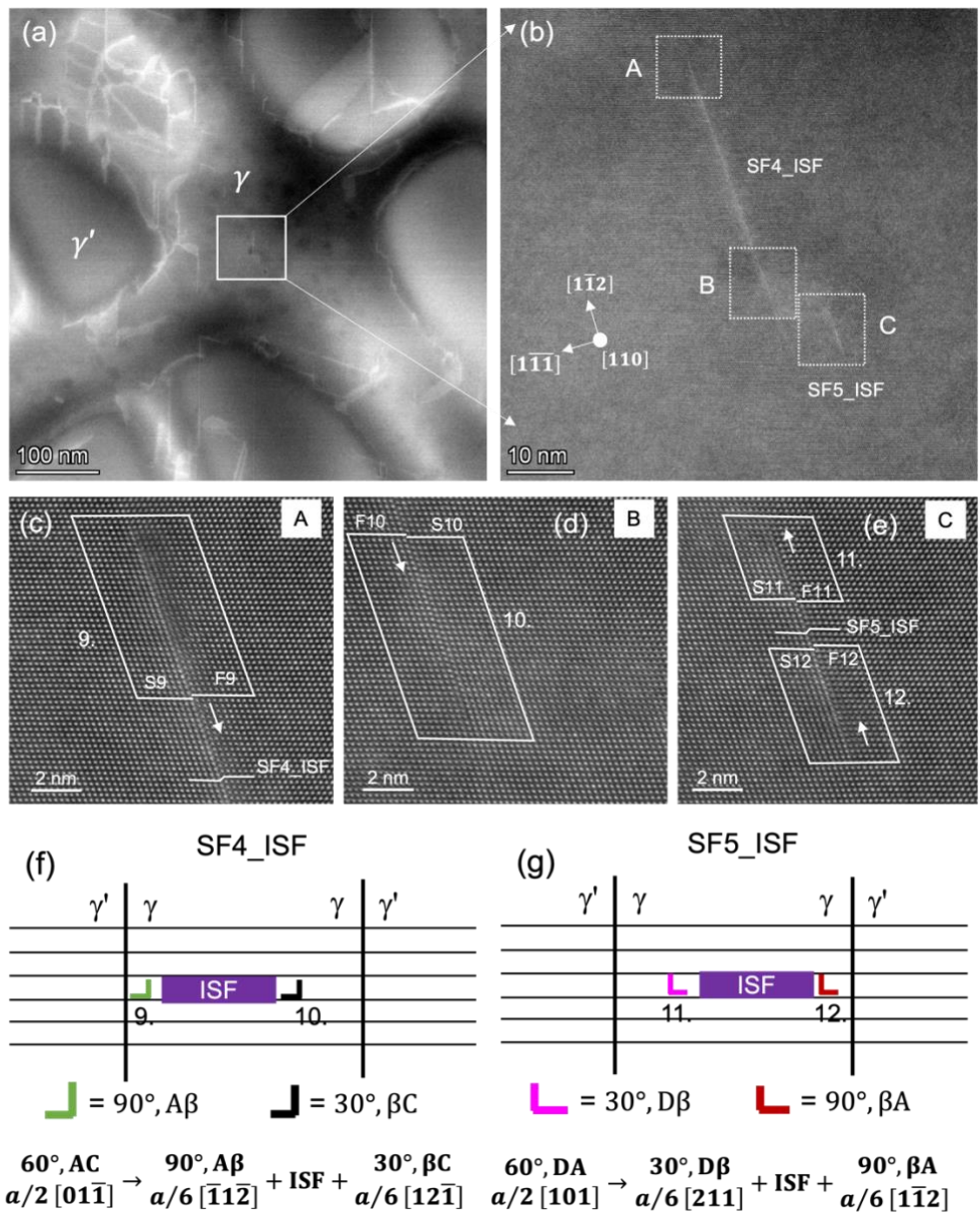

**Fig. 7. Nucleation of ISFs in the γ matrix channel.** (a) HAADF-STEM overview image showing the γ/γ′ two-phase microstructure. (b) Magnified view of the γ matrix channel, revealing two stacking faults (**SF4_ISF** and **SF5_ISF**) and one secondary γ′ precipitate. (c)-(e) Atomic-resolution HAADF-STEM images of the trailing region and leading region of **SF4_ISF**, and overall configuration of **SF5_ISF**, respectively. (f) Schematic illustration of the **SF4_ISF** substructure, comprising a leading Shockley dislocation **(30°, βC)** and one trailing Shockley partial **(90°, Aβ)**, as determined from experimental analysis. (g) Schematic illustration of the **SF5_ISF** substructure, consisting of a leading Shockley dislocation **(90°, βA)** and one trailing Shockley partial **(30°, Dβ)** based on experimental analysis.

## 4.4 The chemical fluctuations at Frank partials of SESF and SISF

Elemental segregation within superlattice SFs and along LPDs in the γ′ phase represents a critical feature influencing fault stability and dislocation mobility. The solutes clustering around LPDs forms Cottrell atmospheres, which exert a solute-drag effect that retards dislocation glide while simultaneously reducing the defect energy. Such segregation can decrease both the stacking-fault and dislocation line energies, thereby facilitating the nucleation of superlattice SFs. To quantify these effects, the local compositions within superlattice SFs and LPDs were analyzed in detail.

**Table 1.** Local chemical compositions (at.%) and corresponding segregation magnitudes for the γ′ phase, leading partial dislocations (LPDs: including Frank and Shockley partials), and fault planes (FPs) associated with SISFs and SESFs, respectively.

| Element | γ′ phase | Frank-type | | | | Shockley-type | | | |
|---|---|---|---|---|---|---|---|---|---|
| | | SISF-FP | SESF-FP | SISF-LPD | SESF-LPD | SISF-FP | SESF-FP | SISF-LPD | SESF-LPD |
| Co | 24.0±0.3 | 29.5±0.1 | 27.0±0.7 | 30.1±0.5 | 32.3±1.9 | 29.4±0.4 | 27.4±0.8 | 30.8±1.2 | 29.4±1.6 |
| Ni | 47.9±0.3 | 42.2±0.2 | 45.1±0.5 | 40.2±0.8 | 38.4±0.6 | 42.8±0.4 | 43.9±0.7 | 41.5±0.1 | 42.1±1.0 |
| Cr | 3.9±0.1 | 5.9±0.1 | 4.8±0.2 | 8.4±0.7 | 9.2±0.6 | 5.5±0.3 | 5.5±0.1 | 6.1±0.1 | 6.4±0.4 |
| Al | 9.5±0.2 | 7.6±0.2 | 7.9±0.2 | 8.5±0.4 | 7.5±1.1 | 7.3±0.7 | 7.8±0.2 | 7.1±0.2 | 8.6±0.9 |
| Ti | 11.4±0.2 | 11.0±0.2 | 11.6±0.6 | 9.7±0.4 | 9.0±0.6 | 11.0±0.2 | 11.7±0.1 | 10.6±0.4 | 10.1±0.0 |
| Mo | 2.2±0.1 | 2.8±0.1 | 2.6±0.9 | 2.5±0.2 | 2.8±0.4 | 3.0±0.1 | 2.7±0.1 | 2.9±0.1 | 2.5±0.2 |
| W | 0.9±0.0 | 1.0±0.0 | 0.92±0.1 | 0.8±0.1 | 0.8±0.1 | 1.0±0.0 | 0.98±0.0 | 1.0±0.1 | 0.8±0.0 |

**Figs. 8(a)** and **9(a)** present HAADF-STEM images of leading Frank partial dislocations associated with SISFs and SESFs, respectively, together with the corresponding elemental distribution maps. In both cases, the fault regions exhibit higher atomic number (Z) contrast than the surrounding γ′ matrix, indicating enrichment of heavier solute elements. The compositional variations along the Frank partials and fault planes display similar segregation patterns for both SISF and SESF configurations.

Elemental mapping reveals clear deviations from the equilibrium γ′ composition, showing enrichment of Co, Cr, Mo, and W, accompanied by depletion of Ni, Al, and Ti at the fault regions. The integrated line profiles across the fault plane (line scan 1) and along the Frank partial dislocation (line scan 2) are shown in **Figs. 8(b)** and **9(b)**, with detailed compositions of the γ′ phase, LPDs, and fault planes (FPs) for SISF and SESF configurations being summarized in **Table 1**. Additional structural and compositional data for Shockley-partial leading configurations are provided in **Supplementary materials 1, Figs. S2 and S3**.

Based on the elemental mapping, Co and Cr are the primary solutes enriched in the Cottrell atmospheres surrounding LPDs. Comparison between line scans 1 and 2 indicates that Co and Cr concentrations at the Frank partial core region (30.1 at.% Co and 8.4 at.% Cr) exceed those at the fault plane (29.5 at.% Co and 5.9 at.% Cr) in the SISF, consistent with the trend observed for the SESF (**Figs. 8(b)** and **9(b)**). Moreover, the SISF fault plane exhibits higher Co and Cr enrichment than that of the SESF (**Fig. 10(a)**), consistent with previous observations [8, 29]. In contrast, the Frank partial of the SISF contains slightly less Co and Cr than the corresponding Frank partial of the SESF (**Fig. 10(b)**). In both SISF and SESF configurations, the Frank partials exhibit greater solute enrichment than the Shockley partials, with Cr showing particularly pronounced accumulation (**Fig. 10(c)**).

This selective segregation primarily originates from the tendency of γ-former elements, particularly Cr and Co, to segregate and locally reduce the CISF energy associated with the leading partial dislocation. Such segregation promotes the accumulation of solute atoms around the dislocation cores, forming a Cottrell atmosphere. On the one hand, the resulting solute-dislocation interactions partially relieve local lattice distortion introduced by LPDs. On the other hand, the Cottrell atmosphere imposes a drag effect that reduces LPD mobility and thereby influencing the extension velocity of SISF and SESFs. These segregation-induced drag effects play a critical role in enhancing the high-temperature deformation resistance of the alloy, as discussed further in **Section 5.4**.

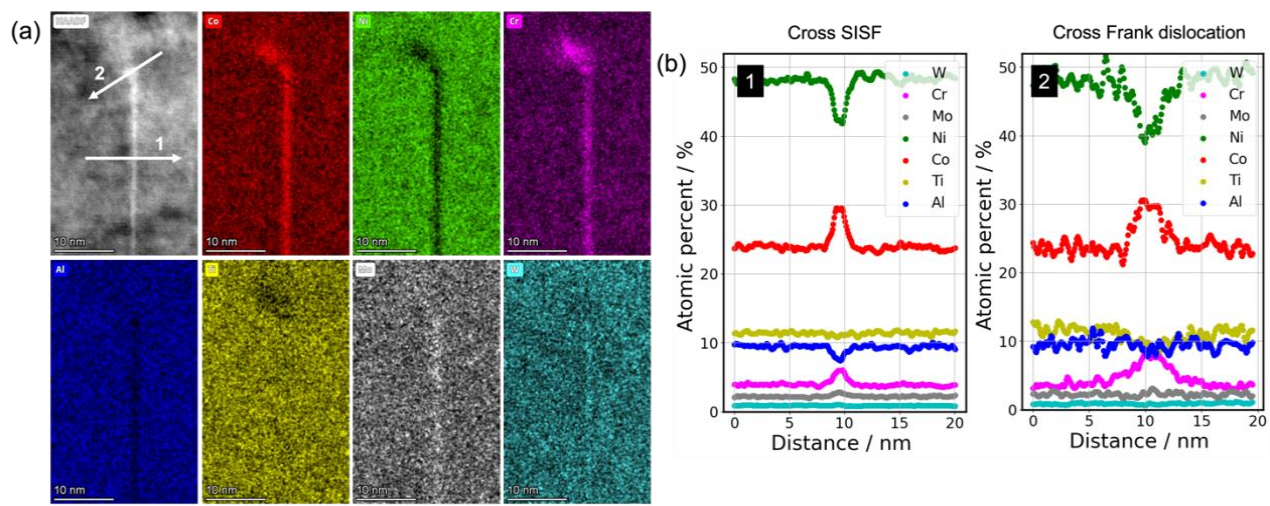

**Fig. 8. The chemical fluctuations at Frank partial dislocation of SISF.** (a) HAADF-STEM image and corresponding STEM-EDXS elemental maps showing enrichment of Co, Cr, Mo, and W, and depletion of Ni and Al along the SF plane. The leading Frank partial is decorated with a Cottrell

atmosphere enriched in Co and Cr. (b) Line profiles of the elemental concentration (Co, Ni, Cr, Al, Ti, Mo, and W) across the SISF (line 1) and the leading Frank partial (line 2).

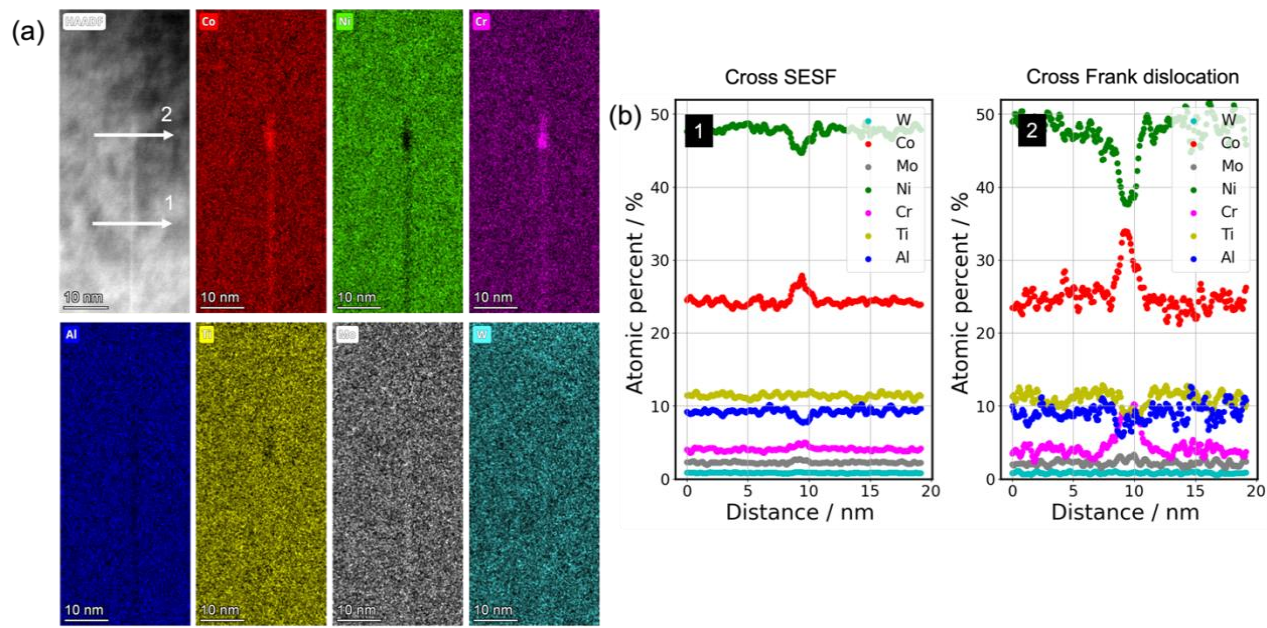

**Fig. 9. The chemical fluctuations at Frank partial dislocation of SESF.** (a) HAADF-STEM image and corresponding STEM-EDXS elemental maps revealing enrichment of Co, Cr, Mo, and W, and depletion of Ni and Al at the SF plane. The leading Frank partial dislocation exhibits a pronounced Cottrell atmosphere with elevated Co and Cr concentration. (b) Line profiles of the alloying elements across the SESF (line 1) and the leading Frank partial (line 2).

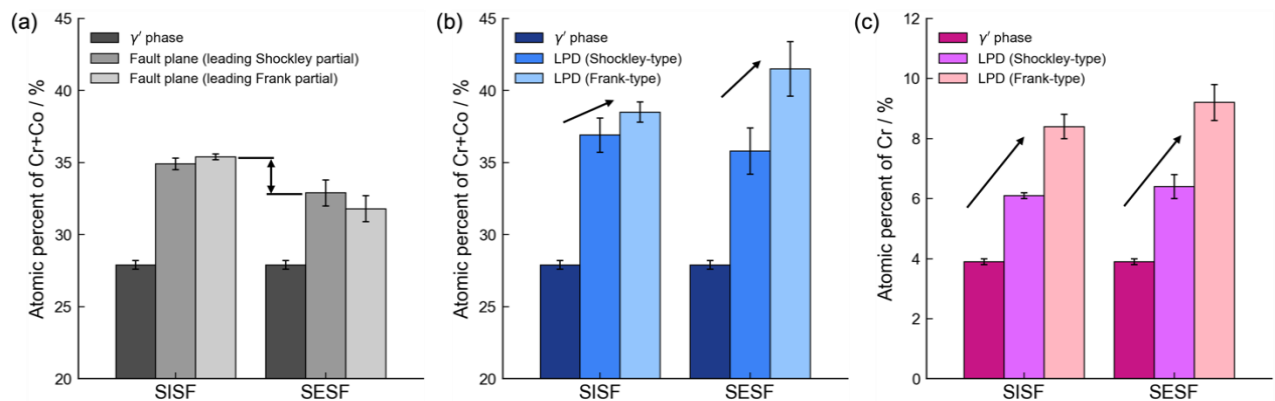

**Fig. 10. Elemental compositions (at.%) within defect regions.** (a) Co + Cr concentration in the γ′ phase and on the fault plane for SISF and SESF configurations with Frank-partial leading and Shockley-partial leading character. The Co + Cr concentration on the fault plane in SISF configuration is higher than in SESF configuration. (b) Co + Cr concentration in the γ′ phase and at the LPD for SISF and SESF with Frank-partial leading and Shockley-partial leading configurations.

(c) Cr concentration in the γ′ phase and at the LPD for the corresponding SISF and SESF configurations.

## 5. Discussion

Based on the above experimental analyses, we identified two new SISF and SESF configurations characterized by Frank leading partials, which differ fundamentally from previously reported SF configurations in both the origin of the LPD and the non-conservative mode of SF expansion, as elaborated in **Section 2**. In contrast to the well-established mechanisms governed by Shockley partial glide, the SISF and SESF configurations identified here arise from the Frank partial climb, introducing a distinct class of SF formation processes in the γ′ phase.

Two types of LPDs, Frank and Shockley partials, are thereby been identified as the carriers shearing the γ′ phase. Their shearing velocities directly influence the deformation rate of the γ′ phase and thus the alloy's creep and high-temperature deformation resistance. To elucidate the underlying mechanisms of our findings, the following key questions are addressed in this section:

(1) What mechanisms govern the nucleation and evolution of the Frank partial-leading SISF and SESF configurations, and how do these differ from the Frank partial-leading SISF pathway reported by Lenz *et al.* [10]?
(2) Are these leading Frank partials bounding ISF and ESF configurations energetically stable prior to its climb into the γ′ phase?
(3) What is the driving force governing both positive and negative climb of Frank partials within the γ′ phase to generate the observed SISF and SESF configurations?
(4) How does the Frank partial climb-assisted formation of SISFs and SESFs compare with the conventional Shockley partial glide-assisted, particularly in the presence of the solute segregation that may alter LPD mobility?

### 5.1. Nucleation of Frank partial dislocations bounding SISF and SESF configurations

Lenz *et al.* [10] demonstrated that the partial dislocations bounding the SISF-3 consist of a trailing 90° edge Shockley partial (**90°, Aβ**) with Burgers vector ($a/6[1\bar{1}2]$) and a leading 90° edge Frank partial (**90°, βB**) with Burgers vector ($a/3[\bar{1}11]$), as illustrated by reaction **Eq. (3)** in **Section 2**. This SISF configuration has been attributed to the splitting of a 90° Lomer dislocation (**90°, AB**). According to Frank's rule, the splitting is strain-energy neutral because the Burgers vectors form a right-angled triangle, such that $|\mathbf{b}_{AB}|^2 + |\mathbf{b}_{A\beta}|^2 = |\mathbf{b}_{\beta B}|^2$. Solute segregation can further reduce the SISF energy, thereby promoting the splitting reaction [10]. Nevertheless, additional energy is still required to overcome the activation barrier for the dissociation of the 90º Lomer dislocation, which explains the relative difficulty of SISF nucleation through this pathway.

In the present work, we identify a more energetically favorable pathway for SISF nucleation, as evidenced in **Fig. 4**. The following subsections provide a detailed analysis of the origin and evolution of this Frank-partial leading SISF configuration, which arises from the dislocation interactions schematically illustrated in **Fig. 11(a)**.

For clarity in the following discussion, the terms **ISF-1**, **ISF-2**, **ISF-3**, and **ISF-4** are used exclusively to distinguish the four ISFs involved in ESF formation in **Fig. 11(a)** and **Fig. 11(b)**, respectively. It should be noted that the labels **SF1_ISF**, **SF2_ISF**, and **SF5_ISF**, used in **Section 4.3**, refer to different ISF configurations observed at the γ/γ′ interface and in the γ matrix channels by HRSTEM.

(1) Before the formation of the Frank partial, a perfect screw dislocation **DC** = $a/2\,[110]$ initiates dissociation into two Shockley partials at the γ/γ′ interface, generating the first ISF (labeled as **ISF-1**), as shown in **Step 1** of **Fig. 11(a)**. The LPD is identified as **βC** = $a/6\,[12\bar{1}]$ (black) and the trailing partial is **Dβ** = $a/6\,[211]$ (magenta). The reaction is:

$$\underset{a/2\,[110]}{0°,\,\mathbf{DC}} \rightarrow \underset{a/6[211]}{30°,\,\mathbf{D\beta}}(\text{trailing}) + \mathbf{ISF}-1 + \underset{a/6[12\bar{1}]}{30°,\,\mathbf{\beta C}}(\text{leading}) \qquad (9)$$

This reaction was directly observed by TEM at the γ/γ′ interface, as evidenced by the **SF1_ISF** region in **Figs. 6(c)** and **(f)**. The reaction is energetically favorable according to Frank's criterion, since the Burgers vectors form a right-angled triangle satisfying $|\mathbf{b}_{\mathbf{DC}}|^2 > |\mathbf{b}_{\mathbf{D\beta}}|^2 + |\mathbf{b}_{\mathbf{\beta C}}|^2$.

(2) In **Step 2**, the leading Shockley partial (**30°, βC**) on the $(\bar{1}11)$ plane intersects with a 60° mixed perfect dislocation (**60°, CB**) (blue) gliding on the conjugate $(1\bar{1}1)$ plane. Their interaction results in the formation of a 90° Frank partial (**90°, βB**) (brown) at the γ/γ′ interface, representing a new synthetic reaction pathway for Frank dislocation generation. The reaction is:

$$\underset{a/6[12\bar{1}]}{30°,\,\mathbf{\beta C}}(\text{leading}) + \underset{a/2\,[\bar{1}01]}{60°,\,\mathbf{CB}} \rightarrow \underset{a/3[\bar{1}11]}{90°,\,\mathbf{\beta B}}(\text{leading}) \qquad (10)$$

This reaction is permissible according to the Burgers vectors satisfy $|\mathbf{b}_{\mathbf{CB}}|^2 + |\mathbf{b}_{\mathbf{\beta C}}|^2 > |\mathbf{b}_{\mathbf{\beta B}}|^2$, indicating a reduction in elastic strain energy upon dissociation and allowing for additional SF generation at the γ/γ′ interface (ISF) or within the γ′ phase (SISF). Although this inequality does not by itself determine the stacking-fault character, it ensures that the dissociation is energetically allowed, thereby enabling the associated intrinsic or superlattice SF to form depending on the local geometry and stacking-fault energy. The resulting fault is formed between a **30°** Shockley partial (**Dβ**) and a **90°** Frank partial (**βB**). This energetically favorable reaction pathway distinguishes from the Frank partial leading SISF mechanism reported by Lenz *et al.* [10], highlighting that Frank partials can be generated through a different class of dislocation interactions at the γ/γ′ interface.

(3) In **Step 3**, the leading Frank partial begins to shear the γ′ precipitate, initiating SISF nucleation via positive climb. This non-conservative climb process is driven by vacancy flux into the leading Frank partial core, facilitating the expansion of the bounding SISF and allowing the Frank partial gradual penetrate into the γ′ phase. The driving forces governing this climb-assisted SISF growth are analyzed in **Section 5.3**.

(4) According to the experimental observations in **Figs. 4(b), (c),** and **(g)**, an ESF is present in the trailing region. We propose that a second ISF (labeled as **ISF-2**) forms on an adjacent $(\bar{1}11)$ plane through the dissociation of a perfect mixed dislocation $\mathbf{AC} = a/2\,[01\bar{1}]$. This dissociation generates a trailing 90° Shockley partial (**Aβ**) (green) and a leading 30° Shockley partial (**βC**) (black), as illustrated in **Step 4** of **Fig. 10(a)**. The reaction is:

$$\underset{a/2\,[01\bar{1}]}{60°,\mathbf{AC}} \rightarrow \underset{a/6[\bar{1}1\bar{2}]}{90°,\mathbf{A\beta}}(\text{trailing}) + \mathbf{ISF-2} + \underset{a/6[12\bar{1}]}{30°,\mathbf{\beta C}}(\text{leading}) \qquad (11)$$

The existence of ISF-2 has been directly confirmed by TEM as **SF2_ISF** in **Figs. 6(b)** and **(e)**. The presence of this second ISF（**ISF-2**）establishes the structural foundation for subsequent ESF formation, providing a necessary intermediate configuration that facilitates the formation for complex local plane fault structures shown in **Fig. 4(g)**.

(5) Finally, **ISF-2** interacts with the trailing segment of **ISF-1**, resulting in the formation of an ESF at the γ/γ′ interface, as illustrated in **Step 5** of **Fig. 11(a)**. As reported by Vorontsov *et al.* [9], a 60° mixed perfect dislocation must climb onto an adjacent {111} plane to interact with another perfect dislocation at the γ/γ′ interface, unless both are coincidentally situated on the same slip plane, a condition that is more likely at elevated temperatures. In the present case, the 60° mixed dislocation (**60°, AC**) may climb onto the adjacent plane of **ISF-1**, where it subsequently dissociates into two Shockley partial dislocations, giving rise to **ISF-2**. The stacking-fault reaction (synthetic interaction) between these two ISFs (**ISF-1** and **ISF-2**) ultimately produces the observed ESF configuration at the γ/γ′ interface. In the FCC structure, a ISF corresponds to the intrinsic stacking sequence ABCAB|ABC, where one stacking layer is missing. When two ISFs occur consecutively, the combined stacking sequence becomes ABCAB|AB|ABC, in which two "C" layers are absent. This pair of intrinsic faults is crystallographically equivalent to an extrinsic fault because the repeated "AB|AB" sequence can be rearranged into the form ABCABC|A|ABCABC, which corresponds to the insertion of an additional "A" stacking layer Therefore, two ISFs can combine to produce a single ESF through the crystallographic equivalence between two intrinsic stacking-fault removals and one extrinsic stacking insertion.

Similarly, **Fig. 11(b)** presents a schematic illustration of the formation mechanism of the Frank-partial leading SESF configuration, for which a representative experimental observation is shown in **Fig. 5**. In this mechanism, the negative climb of a Frank partial leads to the nucleation of SESFs within the γ′ phase. The detailed sequence of reactions involved in SESF formation is outlined in **Steps 1-5** below.

(1) The SESF forms through a mechanism analogous to that of the SISF, utilizing similar "raw materials", namely Shockley partials originating at the γ/γ′ interface or in the γ matrix channel. Prior to the formation of the Frank partial, a perfect pure screw dislocation **DC** = $a/2\,[110]$ begins to dissociate into two Shockley partials, accompanied by the formation of an ISF (labeled as **ISF-3**) at the γ/γ′ interface, as shown in **Step 1** of **Fig. 11(b)**. The leading partial is identified as **βC** = $a/6\,[12\bar{1}]$ (black), and the trailing partial as **Dβ** = $a/6\,[211]$ (magenta). The corresponding reaction equation is identical to reaction **Eq. (9)**.

(2) In addition, a perfect 60° mixed dislocation **DA** = $a/2\,[101]$ begin to dissociate into two Shockley partials, accompanied by the formation of a second ISF (labeled as **ISF-4**) at the γ/γ′ interface, as illustrated in **Step 2** of **Fig. 11(b)**. The leading partial is identified as **βA** = $a/6\,[1\bar{1}2]$ (red) and the trailing partial as **Dβ** = $a/6\,[211]$ (magenta). The corresponding dislocation reaction can be expressed as:

$$\underset{a/2\,[101]}{\overset{60°,\mathbf{DA}}{}} \rightarrow \underset{a/6[211]}{\overset{30°,\mathbf{D\beta}}{}}(\text{trailing}) + \text{ISF} - 4 + \underset{a/6[1\bar{1}2]}{\overset{90°,\mathbf{\beta A}}{}}(\text{leading}) \qquad (12)$$

This reaction was directly confirmed by TEM analysis in the γ matrix channel, as shown for **SF5_ISF** in **Fig. 7(e)** and **(g)**. The reaction is energetically favorable, consistent with the right-angled Burgers vector relationship $|\mathbf{b}_{DA}|^2 > |\mathbf{b}_{D\beta}|^2 + |\mathbf{b}_{\beta A}|^2$.

(3) The **ISF-4** subsequently interacts with the **ISF-3**, leading to the formation of an ESF at the γ/γ′ interface through a stacking-fault reaction, as illustrated in **Step 3** of **Fig. 11(b)**. Based on the dislocation analysis, this process involves the dissociation of one pure screw dislocation and one 60° mixed dislocation, each associated with the generation of two ISFs. The (**30°, βC**) in **Eq. (9)** corresponds to a 30° Shockley partial, while the (**90°, βA**) term in **Eq. (12)** denotes a 90° Shockley partial. The angle between these two Shockley partials (**90°, βA** and **30°, βC**) is 120°, which promotes mutual attraction, resulting in their combination into a (**30°, Dβ**). Consequently, **Eqs. (9)** and **(12)** can be expressed as the following unified reaction:

$$\underset{a/2\,[110]}{\overset{0°,\mathbf{DC}}{}} + \underset{a/2\,[101]}{\overset{60°,\mathbf{DA}}{}} \rightarrow \underset{a/3[211]}{\overset{30°,\mathbf{2D\beta}}{}}(\text{trailing}) + \text{ESF} + \underset{a/6[211]}{\overset{30°,\mathbf{D\beta}}{}}(\text{leading}) \qquad (13)$$

In **Section 4.3**, this dislocation configuration (corresponding to reaction **Eq. (13)**) was directly observed by TEM, although it had already glided into the γ′ precipitate, as shown for **SF3_SESF** in **Fig. 6(g)**.

(4) In **Steps 3-4**, the leading Shockley partial (**30°, Dβ**) on the $(\bar{1}11)$ plane interacts with a mixed 60° perfect dislocation (**60°, BD**) (blue) from the $(1\bar{1}1)$ conjugate plane, resulting in the formation of a Frank partial (**90°, Bβ**) (brown) at the γ/γ′ interface. The reaction can be expressed as:

$$\underset{a/6[211]}{\mathbf{30°, D\beta}}\text{(leading)} + \underset{a/2\,[0\bar{1}\bar{1}]}{\mathbf{60°, CB}} \rightarrow \underset{a/3[1\bar{1}\bar{1}]}{\mathbf{90°, B\beta}}\text{(leading)} \tag{14}$$

This reaction is energetically favorable, consistent with the right-angled Burgers vector relationship $|\mathbf{b_{D\beta}}|^2 + |\mathbf{b_{CB}}|^2 > |\mathbf{b_{B\beta}}|^2$.

(5) In **Step 5**, the leading Frank partial then shears the γ′ precipitate to generate an SESF via a negative climb process. This climb motion requires emission of vacancies from the dislocation core, driven be local chemical potential gradient. The detailed driving forces governing this negative climb and SESF formation are discussed in detail in **Section 5.3**.

In summary, both SISF and SESF formation pathways identified here originates from a previously unrecognized Frank partial generation process: an ISF-leading 60° Shockley partial ($a/6\langle 112\rangle$) interacts with a 60° perfect dislocation ($a/2\langle 110\rangle$) from its conjugate slip plane, producing a Frank partial $a/3\langle 111\rangle$ at the γ/γ′ interface. This reaction is energetically favorable and occurs spontaneously, satisfying the right-angled Burgers vector relationship: $|\mathbf{b}_{a/6\langle 112\rangle}|^2 + |\mathbf{b}_{a/2\langle 110\rangle}|^2 > |\mathbf{b}_{a/3\langle 111\rangle}|^2$. Driven by chemical-potential gradients from the externally applied stress and local defect interactions, the Frank partial undergoes positive climb to generate SISFs or negative climb to generate SESFs. The analytical model describing the climb driving force is presented in **Appendix 3** and discussed in **Section 5.3**.

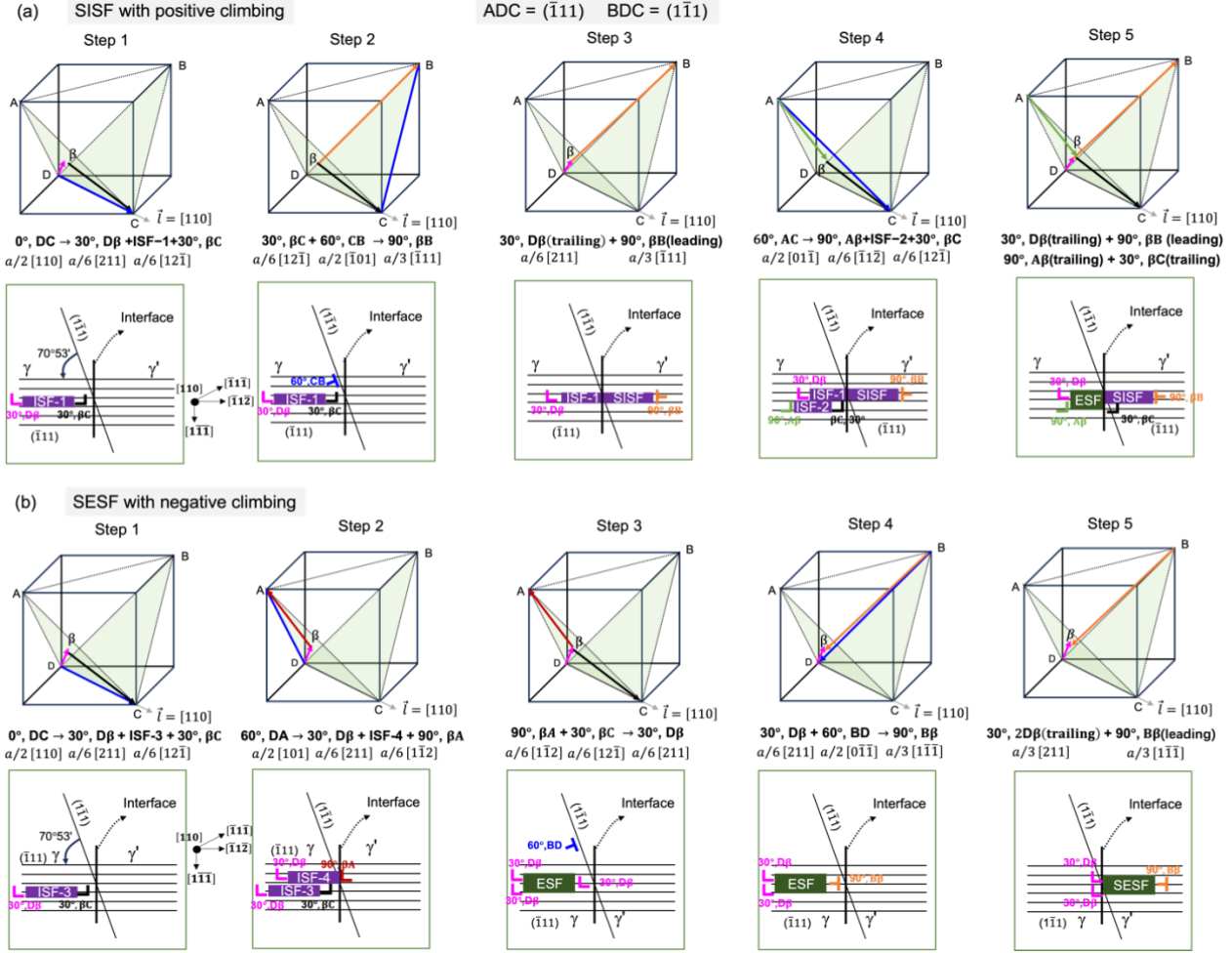

**Fig. 11. Schematic illustration of SISF and SESF nucleation via Frank partial climb.** (a) Formation of a leading Frank partial ($90°, \boldsymbol{\beta B}$) through a synthetic reaction between an ISF-associated leading Shockley partial ($30°, \boldsymbol{\beta C}$) on the $(\bar{1}11)$ plane and a mixed 60° perfect dislocation ($60°, \boldsymbol{CB}$) from the $(1\bar{1}1)$ conjugate plane at the γ/γ′ interface. (b) Formation of an ESF at the γ/γ′ interface through the reaction of two ISFs, followed by the generation of a leading Frank partial ($90°, \boldsymbol{B\beta}$) via the reaction between an ESF-associated leading Shockley partial ($30°, \boldsymbol{D\beta}$) on the $(\bar{1}11)$ plane and a 60° mixed perfect dislocation ($60°, \boldsymbol{BD}$) from the $(1\bar{1}1)$ conjugate plane. The SESFs or SISFs form once the Frank partials climb into the γ′ phase.

## 5.2 The stability of Frank partials in ISF and ESF configurations

According to reactions in **Eqs. (10)** and **(14)**, the Frank partial formation is energetically permissible as the total elastic strain energy decreases upon reaction. Its post-formation stability, however, depends on the elastic interaction between the two dislocations that precedes Frank partial formation. To evaluate whether the newly formed Frank partial will remain stable it is necessary to assess the interaction force between the reacting dislocations. An attractive force between the two reacting dislocations facilitates the formation of a stable Frank partial, and a repulsive force tends to drive its re-dissociation into the original dislocations. For two parallel dislocations, the interaction force per unit length $F$ and the interaction energy per unit length $E$ is given by [37], respectively:

$$F_{i\_j} = \frac{\mu_\gamma}{2\pi r_{i\_j}}(\boldsymbol{b_i} \cdot \boldsymbol{l})(\boldsymbol{b_j} \cdot \boldsymbol{l}) + \frac{\mu_\gamma}{2\pi(1-\nu)r_{i\_j}}[(\boldsymbol{b_i} \times \boldsymbol{l}) \cdot (\boldsymbol{b_j} \times \boldsymbol{l})] \tag{15}$$

$$E_{i\_j} = -\frac{\mu_\gamma(\boldsymbol{b_i}\cdot\boldsymbol{l})(\boldsymbol{b_i}\cdot\boldsymbol{l})}{2\pi}\ln\frac{r_{i\_j}}{r_0} - \frac{\mu_\gamma[(\boldsymbol{b_i}\times\boldsymbol{l})(\boldsymbol{b_i}\times\boldsymbol{l})]}{2\pi(1-\nu)}\ln\frac{r_{i\_j}}{r_0} - \frac{\mu_\gamma[(\boldsymbol{b_i}\times\boldsymbol{l})\cdot\boldsymbol{r_{i\_j}}][(\boldsymbol{b_j}\times\boldsymbol{l})\cdot\boldsymbol{r_{i\_j}}]}{2\pi(1-\nu)r_{i\_j}^2} \tag{16}$$

where $\mu_\gamma$ is the shear modulus of the γ phase (here we use $\mu_\gamma$ = 54.9 GPa at 850 °C [38]), $\boldsymbol{b_i}$ and $\boldsymbol{b_j}$ are the Burgers vectors of the two dislocations, $\boldsymbol{l} = [110]$ is the line direction of the dislocations in this work, $\nu$ = 0.37 is Poisson's ratio and $r_{i\_j}$ is the distance between two parallel dislocations. A negative $F_{i\_j}$ (attraction) stabilizes the Frank partial; a positive $F_{i\_j}$ (repulsion) promotes its separation into constituent partials. $r_0 = b$ is the cut-off radius of a perfect dislocation. If $E_{i\_j}$ decreases as $r_{i\_j}$ decreases, the two-dislocation system becomes more stable in a compact Frank partial configuration, and vice versa.

Based on this concept, a geometric model was established in **Fig. 12(a)** and **(d)** to calculate the interaction forces between dislocations involved in the formation of Frank partials within ISF and ESF configurations.

**Fig. 12(a)** illustrates the triangular geometric relationship between the leading and trailing Shockley partials (**30°, βC** and **30°, βC**) on the primary ($\bar{1}11$) plane, and the 60° mixed perfect dislocation (**60°, CB**) on the conjugate ($1\bar{1}1$) plane. Along the direction of vector $\boldsymbol{r_{2\_1}}$, the resultant interaction force ($F_{total}^{ISF}$) is obtained by summing $F_{2-3}cos\alpha$ and $F_{2-1}$, with the analytical expressions provided in **Appendix 2, Eqs. (A.2)** and **(A.3)**. The total interaction energy ($E_{total}^{ISF}$) is the sum of $E_{2-3}$ and $E_{2-1}$, based on **Appendix 2, Eq. (A.4)**. The total interaction force shown in **Fig. 12(b)** reveals that, within the given configuration, the leading partial dislocation 1 (**30°, βC**) attracts dislocation 2 (**60°, CB**). Consistently, the total interaction energy decreases from positive to negative as dislocation 2 approaches the leading partial 1, as shown in **Fig. 12(c),** demonstrating that the formation of a Frank partial is more stable than remaining as dislocation (**60°, CB**) and a leading partial (**30°, βC**). Consequently, the dislocation (**60°, CB**) reacts with leading partial (**30°, βC**) to form a stable Frank partial (**90°, βB**).

A similar triangular geometric configuration exists for the ESF, as illustrated in **Fig. 12(d)**. Using **Eq. (15)**, the interaction forces $F_{5-4}$ and $F_{5-6}$ between (**30°, Dβ**) – (**60°, BD**) and (**30°, 2Dβ**) – (**60°, BD**) were evaluated as functions of $r_{5-4}$. Along the direction of vector $\boldsymbol{r_{5\_4}}$, the resultant interaction force ($F_{total}^{ESF}$) is obtained by summing $F_{5-6}cos\alpha$ and $F_{5-4}$, with the analytical expressions provided in **Appendix 2, Eqs. (A.6)** and **(A.7)**. The total interaction energy ($E_{\text{total}}^{\text{ESF}}$) is the sum of $E_{5-6}$ and $E_{5-4}$, based on **Appendix 2, Eq. (A.8)**.

The calculated results indicate that both the leading and trailing partials exert attractive forces on the perfect dislocation, though the magnitude of attraction varies with the inter-dislocation spacing. As shown in **Fig. 12(e)**, the interaction force between the dislocation (**60°, BD**) and the leading partial (**30°, Dβ**) becomes progressively more negative as the distance $r_{5-4}$ decrease, signifying a rapidly strong attraction. This strong interaction drives the reaction between the perfect dislocation (**60°, BD**) and (**30°, Dβ**), giving rise to a stable Frank partial (**90°, Bβ**). Moreover, **Fig. 12(f)** demonstrates that the interaction energy $E_{\text{total}}^{\text{ESF}}$ becomes negative as the perfect dislocation (**60°, BD**) approaches the leading partial (**30°, Dβ**). This trend confirms that the attractive interaction strengthens continuously with decreasing $r_{5-4}$, promoting the formation and stability of the Frank partial.

Based on the analysis of the interaction forces and energies presented above, the proposed pathway for Frank partial formation in both ISF and ESF configurations (**Section 5.1**) is energetically favorable.

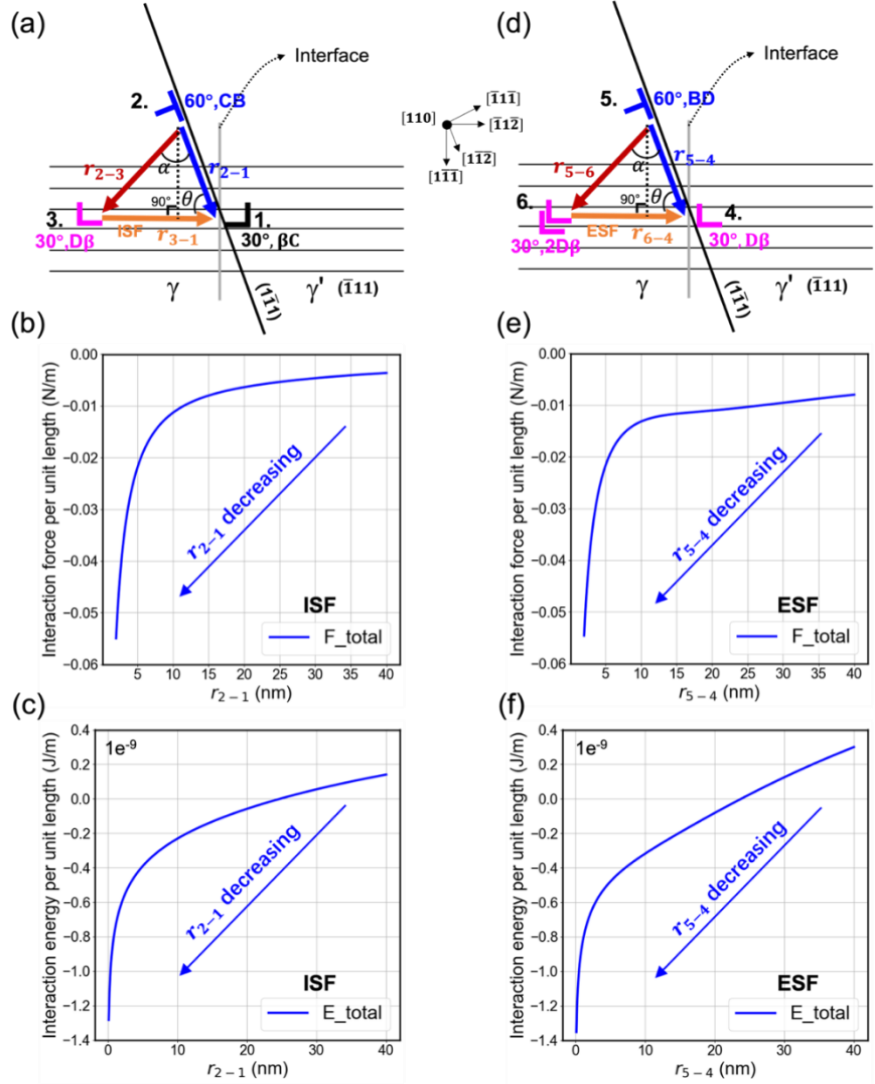

**Fig. 12** (a), (b) and (c) Geometrical model and corresponding interaction force, energy between the leading and trailing Shockley partial ($30°, \boldsymbol{\beta C}$ and $30°, \boldsymbol{\beta C}$) on the ($\bar{1}11$) plane and the 60° mixed perfect dislocation ($60°, \boldsymbol{CB}$) on the conjugate ($1\bar{1}1$) plane in the ISF configuration. (d), (e) and (f) Geometrical model and corresponding interaction force, energy between the leading and trailing Shockley partials ($30°, \boldsymbol{D\beta}$ and $30°, \boldsymbol{2D\beta}$) on the ($\bar{1}11$) plane and the 60° mixed perfect dislocation ($60°, \boldsymbol{CB}$) on the conjugate ($1\bar{1}1$) plane in the ESF configuration.

## 5.3 Climb of Frank partial in the γ′ phase

Following the discussion of the origin and the stability of the Frank partials in **Section 5.1** and **5.2**, respectively, we now examine the driving forces that enable Frank partials to climb into the γ′ phase and assess the associated climb rate.

### 5.3.1 Driving force of Frank partial climb

Dislocation climb is a diffusion-controlled, non-conservative process that allows the motion of dislocations out of its original slip plane by absorbing or emitting point defects (vacancies or interstitials). The configurational forces governing this process can be grouped into three types [39]: i) the climb component of the total Peach-Koehler (PK) force, $f_{cl}$. ii) the osmotic force, $f_{os}$, arising from the change of free energy due to the creation or annihilation of vacancies at dislocation core during climb. iii) the drag force, $f_{dr}$, which includes both the vacancy free-energy change associated with the evolving stress field due to climb, and specific to the present configuration (a SF bounding a leading Frank partial and a trailing Shockley partial), an additional resistance from the energy cost of widening the superlattice SF. A full derivation of the total climb driving force is presented in **Appendix 3**; here we summarize only the essential physics and governing equations relevant to climb-mediated SISF and SESF formation.

i) The climb component of the PK force, $f_{cl}$, contains contributions from the externally applied stress $\sigma_{app}$, the elastic interactions with surrounding defects $\sigma_{inter}$, and the dislocation self-stress $\sigma_{self}$,

$$f_{cl} = ((\boldsymbol{\sigma}_{app} + \boldsymbol{\sigma}_{inter} + \boldsymbol{\sigma}_{self}) \cdot \boldsymbol{b}_F) \times \boldsymbol{l}_F \cdot \boldsymbol{n} \tag{17}$$

where $\boldsymbol{b}_F = a/3[1\bar{1}\bar{1}]$ and $\boldsymbol{l}_F = [110]/\sqrt{2}$ are the Burgers vector and the normalized line vector of the Frank partial, respectively. $\boldsymbol{n} = \frac{\boldsymbol{b}_F \times \boldsymbol{l}_F}{|\boldsymbol{b}_F \times \boldsymbol{l}_F|}$ is the climb direction. For the isolated dislocation pair considered here, as discussed in **Appendix 3**, the elastic interaction between the leading Frank partial and the interface-pinned Shockley partial is zero for the present geometry. The self-stress of an edge dislocation is anti-symmetric, therefore it only changes the local vacancy redistribution rather than a net vacancy flux in the radially symmetric diffusion model adopted here. Hence, $f_{cl}$ is dominated by the externally applied stress and can be estimated as,

$$f_{cl} = \frac{1}{3}(\sigma_{33} + \sigma_{22} + \sigma_{11})b_F \tag{18}$$

Note that this estimation provides a representative magnitude of the contribution from the externally applied stress. The actual stress acting on a given Frank partial may vary with the corresponding local stress partitioning in the γ′ phase.

ii) The osmotic force, $f_{os}$, can be determined by the vacancy concentration difference between the dislocation core $c_v^c$ and the far field $c_v^0$:

$$f_{os} = -\frac{kTb_F}{\Omega_f} \ln\left(\frac{c_v^c}{c_v^0}\right) \tag{19}$$

where $\Omega_f$ is the vacancy formation volume.

iii) The drag force, $f_{dr}$, consists a small contribution from the vacancy free-energy change with the evolving dislocation stress field due to climb [38], and a dominant term from the change in SF energy as the partial climbs. By neglecting the former, the drag force per unit length of dislocation can be expressed as

$$f_{dr} = \gamma_{SF} \tag{20}$$

which always resist the expansion of the SF. It is therefore positive for SESF case, as it resists the negative climb induced SF expansion, and negative for SISF case, as it resists positive climb. The magnitudes of these three forces in **Eqs (18)-(20)** are evaluated in the following subsections.

### 5.3.2 Climb rate of the Frank partial

Vacancy diffusion along the dislocation line is rapid due to pipe diffusion, allowing the dislocation core maintain a near-equilibrium vacancy concentration, $c_v^c$. In contrast, vacancy diffusion in the surrounding lattice is governed by much slower bulk diffusion, which establishes a concentration gradient between the dislocation core and the far field. For an isolated dislocation segment in an otherwise homogenous medium, mass conversation and cylindrical symmetry naturally give rise to dislocation climb driving by a radial vacancy flux, as adopted in **Appendix 3**. Although the presence of SFs breaks perfect cylindrical symmetry of leading Frank partial, these geometric deviations do not alter the essential coupling between vacancy flux and climb, and the radial flux model provides a suitable first-order approximation for evaluating climb rates.

By combining the radial diffusion field with the configurational forces defined in **Eqs. (18)-(20)**, the Frank partial climb velocity is derived as (**Eq. (A.24)** in **Appendix 3**),

$$v_c = \frac{j_r 2\pi r}{b_F} = \frac{2\pi D_v}{b_F \ln(r_0/r_c)} c_v^0 \left[1 - \frac{c_v^c}{c_v^0}\right] \tag{21}$$

where the vacancy concentration at dislocation core is given by,

$$c_v^c = c_v^{eq} = c_v^0 \exp\left(\frac{(f_{cl} + f_{dr})\Omega_f}{b_F kT}\right) \tag{22}$$

**Eqs. (21)-(22)** highlight two key factors controlling Frank partial climb: i) a kinetic perfector, $\frac{2\pi D_v}{b \ln(r_0/r_c)}$, governed by the vacancy diffusivity, reflecting how fast the transport capability of vacancies in the material. It is highly temperature-dependent, as $D_v = D_0 \exp\left(-\frac{E_m}{kT}\right)$, with $D_0$ representing the diffusivity pre-factor and $E_m$ representing the vacancy migration energy. The geometric factor $\ln(r_0/r_c)$ is taken as 4.6, corresponding to an assumed ratio of $r_0/r_c$ = 100, which is typical for dislocation climb analyses in metals and alloys [40]. ii) an exponential term, $c_v^0 \left[1 - \exp\left(\frac{(f_{cl}+f_{dr})\Omega_f}{b_F kT}\right)\right]$, capturing the magnitude and direction of the net vacancy flux driving the climb motion, where $c_v^0 = exp\left(-\frac{E_f}{kT}\right)$ represents the reference vacancy concentration at far-field under zero-stress state, and $E_f$ representing the vacancy formation energy. In the present study, an applied compressive stress of 572 MPa corresponds to $\sigma_{33} = -572$ MPa. Collectively, these two factors, kinetic transport and net vacancy driving force, quantify how material properties, geometry, stress, and temperature synergistically control the Frank partial climb, with remaining parameters summarized in **Table 2**.

**Eq. (21)** demonstrates that the Frank partial climb rate is sensitive to the vacancy diffusion coefficient and therefore to alloy compositions. For example, refractory transition metals such as Ta, W, Re, and Nb are known to increase both the vacancy formation energy and diffusion barrier in Co- and Ni-based alloys [41, 42], thereby reducing vacancy concentration and mobility. As a result, the addition of these refractory elements in next-generation CoNi-based superalloys is expected to supress vacancy-mediated climb process.

### 5.3.3 The effect of solute segregation on climb rates

We now examine the effect of solute segregation on the climb process. Solutes segregation alters the local chemical composition at both the fault plane and the dislocation core. As shown in **Fig. 8** and **9**, significant enrichment of Cr and Co occurs along the SF plane and at the leading Frank partial core. Such segregation strongly reduces the effective SF energy, $\gamma_{SF}$ [30], thereby lowering the energetic penalty associated with fault widening. According to previous work [29], $\gamma_{SF}$ decreases from 89 mJ/m² to 18 mJ/m² for SESF and from 68 mJ/m² to 10 mJ/m² for SISF upon segregation, significantly promoting fault expansion and thus facilitating climb of the bounding Frank partial.

In contrast, solute segregation at the dislocation core tends to stabilize the core structure by slightly relieving the local strain energy [43]. This stabilization reduces both the vacancy relaxation volume $\Delta\Omega$ and the local hydrostatic stress in the immediate core region. As a result, the stress induced vacancy formation/annihilation are weakened, producing a smaller vacancy concentration gradient to drive climb. Atomistic and thermodynamic analysis [43, 44] confirm that such screening of hydrostatic stress field does occur in substitutional solid solutions such as Ni-Al, however, the effect is weak, and the resulting stress relief is confined to a very small region around the dislocation core and remains limited in magnitude. Moreover, the solute-vacancy binding energies for Cr and Co in Ni are very small ($|E_b| \lesssim 0.05$ eV) [45], and thereby does not lead to vacancy trapping or any significant modification of vacancy mobility. Our estimation of the volume change within the Cottrell atmosphere indicates that it differs from that of the surrounding γ′ lattice by only ~ 4% (**Supplementary Materials 2**), further supporting the conclusion that the local stress modification is minor. Therefore, the reduction in dislocation climb rate associated by solute-decorated core is expected to be minor, being consistent with Mishin *et al.* conclusion [43]. Overall, the dominant influence of solute segregation in the present system arises from the substantial reduction of $\gamma_{SF}$ at the fault plane, whereas the secondary effects associated with leading partial dislocation core stabilization and stress relief are comparatively negligible under the conditions of this study.

We next evaluate the climb velocity for both SESF and SISF configurations, with and without solute segregation. For a SISF configuration, the drag force per unit length caused by SF is $f_{dr} = \gamma_{SISF}$, yielding a climb rate of – 0.34 nm/s. Whereas in the presence of solute segregation, the SF energy is reduced, such that the drag force becomes $f_{dr} = \gamma_{SISF} - \Delta\gamma_{SISF}$, and the corresponding climb rate changes to + 0.32 nm/s. Thus, in the absence of solute segregation, the SISF shrinks due to a negative climb rate. In contrast, solute-induced reduction of the SF energy reverses the driving force, resulting in a positive climb rate and thus the expansion of the SISF. It indicates that the solute segregation contributes to the main driving force facilitating the Frank partial climb to expand the SF; without it, the SISF is predicted to shrink, highlighting the critical role of solute segregation in the formation of SISFs.

For the SESF configuration, the drag force is $f_{dr} = -\gamma_{SESF}$ in the absence of segregation and $f_{dr} = -(\gamma_{SESF} - \Delta\gamma_{SESF})$ when segregation is present. The corresponding climb velocities are + 1.24 nm/s without segregation, and + 0.61 nm/s with segregation. Following our sign convention, a positive climb rate corresponds to SESF shrinkage. The simplified climb model therefore predicts that an existing SESF tends to shrink under the applied macroscopic compressive stress, in contrast to the experimentally observed formation of SESFs via negative Frank climb.

This apparent discrepancy arises from the assumptions adopted in the analytical model, which assumes an isolated Frank-Shockley pair bounding an SESF, subjected to a uniform far-field stress and with spatially homogenous SF energy. Under such idealized conditions, the sign of $f_{cl} + f_{dr}$ reflects only the incremental thermodynamic tendency of an existing SESF segments to shrink or expand. In reality, however, SESF nucleation occurs in a more complex local environment, where the Frank partial is generated by reactions involving γ/γ′ interface dislocations and Shockley partials. In these regions, the local resolved stress from defect interactions in the climb direction can exceed the applied macroscopic stress, and the effective PK climb force may differ substantially in both magnitude and sign from the value used in the current analytical model.

**Table 2.** Materials parameters adopted for climb and glide velocity calculations of leading partial dislocations (Shockley and Frank partials) in the γ′ phase.

| Parameters | Values |
|---|---|
| Temperature, $T$ | 850 °C |
| Shear modulus, $\mu_{\gamma'}$ | $58.6\ GPa$ of the γ′ phase at 850 °C [38] |
| Poisson's ratio, $\nu$ | 0.37 |
| Boltzmann's constant, $k$ | $1.380649 * 10^{-23}\ J/K$ |
| Burgers vector of Frank partial $a/3[1\bar{1}\bar{1}]$, $b_F$ | 0.206 nm |
| Burgers vector of Shockley partial $a/6[211]$, $b_S$ | 0.146 nm |
| Applied stress, $\sigma$ | $-572\ MPa$ (yield stress for CoNi-2Mo1W) |
| vacancy formation energy, $E_f$ | 164 kJ/mol [41] |
| Vacancy migration energy, $E_m$ | 120 kJ/mol [41] |
| Diffusivity pre-factor, $D_0$ | $4 \times 10^{-6}\ m^2/s$ |
| Solute composition in Cottrell atmosphere of SISF, $c_{solute}$ | $c_{solute} = 4.88 \times 10^{27}$ m$^{-3}$ for 6.1 at.% Cr |
| Solute composition in Cottrell atmosphere of SESF, $c_{solute}$ | $c_{solute} = 5.12 \times 10^{27}$ m$^{-3}$ for 6.4 at.% Cr |
| Diffusivity of solute Cr atoms, $D_{solute}$ | $D_{Cr} = 12.83 \times 10^{-18}\ m^2/s$ [50] |
| The factor related to the interaction energy between the dislocation and the solutes, $\beta$ | $\beta = \frac{\mu_{\gamma'} b_S}{3\pi} \frac{1+\nu}{1-\nu}(0.1 b_S^3)$ [37] |
| SESF energy, $\gamma_{SESF}$ | $89\ mJ/m^2$ [46] |
| Energy reduction for SESF (solutes segregation), $\Delta\gamma_{SESF}$ | $-71\ mJ/m^2$ [29] |
| SISF energy, $\gamma_{SISF}$ | $68\ mJ/m^2$ [46] |
| Energy reduction for SISF (solutes segregation), $\Delta\gamma_{SESF}$ | $-58\ mJ/m^2$ [29] |

### 5.3.4 Effect of local defect interactions on Frank partial climb

To illustrate the potential role of local defect interactions, we evaluate the stress field of a simplified γ/γ′ microstructure containing a pile-up of three infinitely long, straight edge dislocations located at the γ/γ′ interface (**Supplementary Material 1, Fig. S4**). The total stress field is estimated by superimposing the analytical elastic solution for dislocation pile-up onto a two-dimensional finite-element (FE) solution that accounts for both the γ/γ′ lattice misfit and an externally applied uniaxial compressive load of − 572 MPa. In this simplified model, the interfacial dislocations are assumed to be evenly spaced, with a separation of 30 nm. The FEM calculation captures the combined contribution of lattice misfit stresses and the applied macroscopic loading, while the analytical dislocation field provides an estimate of the additional stress arising from interfacial dislocation pile-ups. Although the model adopts several simplifying assumptions and does not represent actual experimental configurations, it demonstrates that plausible defect configurations can generate substantial local stress concentrations. In particular, the local stress within the γ′ phase near the γ/γ′ interface becomes positive and reaches values on the order of + 200 MPa, significantly different from the uniform stress state assumed in the climb model. These local stress concentrations can promote vacancy supersaturations near the Frank partial dislocation core, *i.e.* $\frac{c_v^c}{c_v^0} > 1$, thereby driving the Frank partial to absorb vacancies and undergo negative climb to widen the SESF.

Thus, the experimentally observed SESF formation by negative climb is compatible with a microstructural environment dominated by localized defect interactions and chemical-potential gradients that are not captured in the simplified analytical formulation. The analytical model should therefore be viewed as a framework for assessing the relative contribution of climb to high-temperature deformation, while the actual climb behavior is governed by complex local defect interactions. Nevertheless, the calculated climb velocities for the SISF and SESF configurations, summarized in **Table 3**, provide baseline, order-of-magnitude reference values derived under uniform far-field stress and homogeneous stacking-fault energy.

**Table 3.** The calculated Frank climb and Shockley glide velocity in the case of SISF and SESF under solutes segregation. (Minus - means the negative climb; Plus + means the positive climb.)

|  | SISF | SISF (with solute segregation) | SESF | SESF (with solute segregation) |
|---|---|---|---|---|
| Defects energy | 68 mJ/m$^2$ | 10 mJ/m$^2$ | 89 mJ/m$^2$ | 18 mJ/m$^2$ |
| Frank climb velocity | − 0.34 nm/s | + 0.32 nm/s | / | / |
| Shockley glide velocity | / | 1.20 nm/s | / | 0.94 nm/s |

### 5.4 Glide of Shockley partial in the γ′ phase

To compare the climb- and glide-assisted SFs, we further evaluate the glide rate of Shockley partial bounding similar SISF and SESF configurations. Strong segregation of solute elements such as Co and Cr occurs at the leading Shockley dislocations, forming a Cottrell atmosphere, as shown in **Fig. 10(b)**. The Cottrell atmosphere associated with the leading Shockley dislocation is believed to develop through pipe diffusion from the γ matrix. The velocity of the Cottrell atmosphere around Shockley dislocations can be estimated using the approach proposed by Titus *et al.* [47] and other researchers [48, 49].

The motion of a dislocation interacting with diffusing solute atoms is strongly influenced by the redistribution of solutes around its core. Depending on the relative rates of dislocation glide and solute diffusion, two kinetic regimes can be distinguished. In the fast regime, the dislocation moves too rapidly for the solute atmosphere to remain attached, achieving near-sonic velocities. In contrast, in the slow regime, the dislocation drags an incompletely developed solute cloud, experiencing a viscous drag force that limits its glide velocity. Wu *et al.* [50] reported that higher creep strain rates correspond to a lower solute concentration in the Cottrell atmosphere. This solute-drag-controlled mechanism, first proposed by Cottrell *et al.* [51], explains the deformation behavior of metals under low-stress, high-temperature conditions where solute diffusion is rate-controlling.

In this work, due to high solute concentration (Co and Cr) around leading Shockley partials, we adopt the "slow dislocation" assumption, treating glide as solute-drag-limited motion. The dislocation glide velocity is determined by the balance between the applied driving stress and the diffusional relaxation of the surrounding solute field. The shearing velocity of the leading Shockley partials, dragged by Cottrell atmosphere, can be estimated using the equation [37]:

$$v_g = \frac{D_{solute} kT}{\beta^2 c_{solute} \ln\left(\frac{r}{r_0}\right)} \left(\tau - \frac{\gamma_{SF}}{b_S}\right)(2b_S) \tag{25}$$

where, $\tau = \sigma_Y/M$ is the relative shear stress on the Cottrell atmosphere, M = 3.06 is the mean orientation factor for the FCC polycrystalline matrix [52], and $\gamma_{SF}$ is the superlattice SFs energy. The force per unit length on two Shockley partials ($a/3[12\bar{1}]$) in SESF or one Shockley superpartial ($a/3[12\bar{1}]$) in SISF is $\tau(2b_S) = 4.08 \times 10^{-2}\ N/m$. The other parameters ($D_{solute}, c_{solute}, \beta, k, T, \sigma_Y, b_S, \ln(r/r_0)$) are shown in **Table 2**.

The glide velocity of Shockley partials is particularly sensitive to the diffusion rate of solute atoms segregating toward the dislocation core. In present Co-based alloy, W is the slowest diffusing element, followed by Cr [42]. However, W segregation does not contribute significantly to the formation of the Cottrell atmosphere. Instead, Cr segregation plays a dominant role in facilitating solute cloud migration during shearing. Under constant applied stress and temperature, the Shockley dislocation glide velocity strongly depends on the Cr concentration and diffusivity in the Cottrell atmosphere. The migration velocity of the Cottrell atmosphere associated with Shockley dislocation was estimated based on Cr diffusion coefficient.

The calculated glide velocity of Cottrell atmosphere associated with the leading Shockley dislocations in SISF and SESF configurations is approximately 1.42 nm/s and 1.11 nm/s, respectively. As shown in **Table 3**, the glide velocity of the Shockley partial with a Cottrell atmosphere is comparable to the Frank partial climb velocity. Based on analysis above, one efficient method to suppress Cottrell atmosphere migration is decreasing solute diffusivity.

In both Ni and Co, the refractory element Re has the lowest diffusion coefficient among transition metals [1, 41, 42]. For example, in Ni-based superalloys, such as ERBO 1 and CMSX-4, the refractory element Re, followed by Co and Cr, exerts the most pronounced effect on the Cottrell atmosphere, the so-called "Re effect". The Cottrell atmosphere with Re segregation surrounding the Shockley dislocations within the γ′ phase impedes their glide and lowers its shear strain rate at high temperatures.

In addition, the addition of W, Ta and Nb can increase the shear modulus of the γ′ phase [53], which enhances the interaction factor (***β***) related to the dislocation-solute binding energy (**Eq. (25)** and **Table 2**) and further reduces the glide velocity of the leading Shockley dislocation.

Based on analysis in **Section 5.3** and **5.4**, alloying with Re, W, Nb, and Ta in next-generation CoNi-based superalloys will represent an effective strategy to suppress the mobility of Shockley dislocation glide with Cottrell atmospheres and Frank partial climb. In addition, the group-V elements W, Ta and Nb are known to segregate into SISFs and SESFs, stabilizing $D0_{19}$ and $D0_{24}$ ordered structures [15, 22, 24, 47, 54-56], respectively. These local phase transformation within SESFs can further suppress the thickening and lengthening of SESFs into microtwins, thereby enhancing the microstructural stability and high temperature deformation resistance of superalloys.

## 6. Conclusion

In this work, a new mechanism for SISF and SESF formation in CoNi-based superalloys was identified, involving the positive and negative climb of a Frank partial dislocation ($a/3\langle 111 \rangle$), respectively. Notably, the nucleation of SESF via the negative climb of a Frank partial was firstly confirmed by experiment. The main conclusions are as follows:

(1) A more energetically favorable pathway for SISF nucleation via positive Frank partial climb is proposed, differing from the mechanism suggested by Lenz *et al.* [10]. We believe that the Frank partial formed at the γ/γ' interface is via synergy of one leading 30° Shockley partial with ISF and one 60° mixed perfect dislocation from its conjugate {111} planes, rather than Lomer-type dislocation splitting. Despite these findings, the existence of SISFs nucleation mechanism via splitting of the Lomer dislocation in a pair of Shockley partial and Frank partial cannot be ruled out.

(2) Similarly, SESF nucleation occurs via the negative climb of a Frank partial formed by the interaction of a 30° Shockley partial with an ESF and a 60° mixed perfect dislocation from its conjugate {111} plane.

(3) The stability of the Frank partial formed through the proposed new pathway was evaluated, indicating that it is energetically difficult to dissociate. This stability promotes SISF and SESF nucleation within the γ′ phase through Frank partial climb across the complex γ/γ' interface.

(4) An analytical climb model was developed to quantify how the macroscopically applied stress, together with local defect interactions, and the drag force from the bounding stacking faults, modifies the vacancy concentration gradient at the Frank partial core. These variations in chemical potential govern vacancy absorption or emission during deformation, thereby driving the non-conservative climb of the Frank partial and enabling the associated expansion of superlattice stacking faults.

(5) The leading dislocations mobility comparison reveals that Shockley partial glide velocity, under effect of Cottrell atmosphere, is comparable to the Frank partial climb velocity during high-temperature deformation in the γ′ precipitates.

By identifying Frank-partial climb as an essential yet previously overlooked nucleation pathway for both SISFs and SESFs, this study establishes a new mechanistic basis for planar-fault formation in γ′-strengthened superalloys. This insight not only clarifies the origin of complex fault configurations formed under high-temperature deformation, but also provides new opportunities for understanding planar-fault shearing mechanism of superalloys at high-temperatures. Future work combining multiscale modeling with in situ high-temperature experiments is needed to quantitatively determine the role of partial dislocation climb in creep and its coupling with solute segregation and vacancy-mediated transport.

**Acknowledgements**

Dr. Zhida Liang acknowledges the support of transmission electron microscopy (TEM, Talos 200i) facilities provided by Prof. Florian Pyczak's group at Helmholtz-Zentrum Hereon GmbH, Germany.

**Declaration of generative AI and AI-assisted technologies in the writing process**

During the preparation of this work the author(s) used ChatGPT in order to improve writing. After using this tool/service, the author(s) reviewed and edited the content as needed and take(s) full responsibility for the content of the publication.

**Appendix 1. The summary of SISF nucleation via $a/3\langle 112 \rangle$ superdislocation shearing**

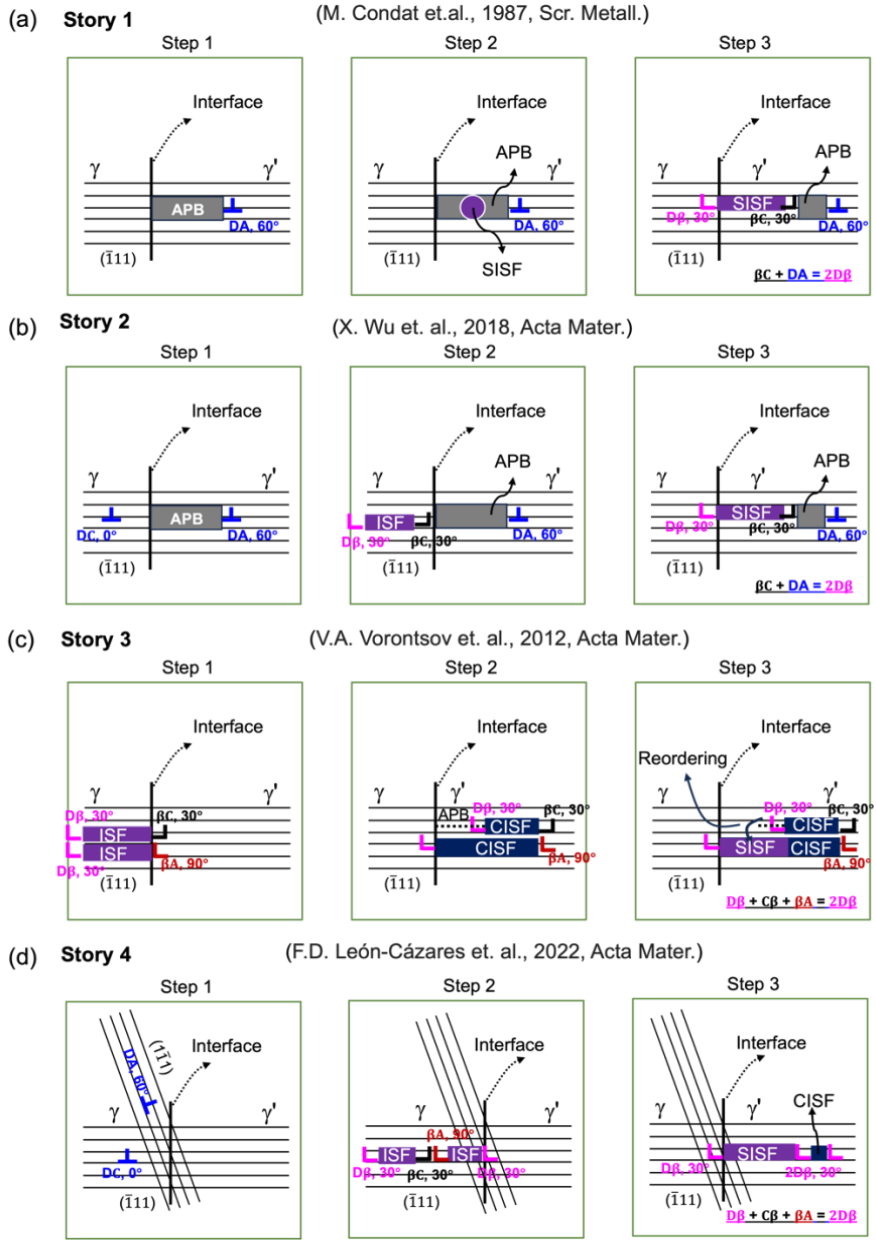

**Fig. A.1. The summary of SISF nucleation via Kear mechanism:** (a) In **Story 1**, Condat and Decamps [19] (1987) proposed that a single $a/2\langle 110\rangle$ $\gamma$ dislocation penetrates the $\gamma'$ phase, producing a planar defect of APB. Subsequently, one $a/6\langle 112\rangle$ Shockley partial loop nucleates, as is shown in **Fig. A.1(a)**. One segment of this loop reacts with the leading $a/2\langle 110\rangle$ dislocation to form a $a/3\langle 112\rangle$ superpartial, while the remaining segment moves back to the $\gamma/\gamma'$ interface, transforming the high-energy APB into a low-energy SISF. This mechanism has recently been validated by molecular dynamics simulations [21]. (b) In **Story 2**, Wu *et al.* [20] (2018) proposed that a 60° channel dislocation approaches another 60° interface dislocation with a distinct Burgers vector on the same crystallographic plane. The second 60° perfect dislocation dissociates into two

Shockley partials, which pushes the first 60° dislocation into the γ′ phase, creating a small APB. The leading Shockley partial then reacts with the first dislocation to form an overall $a/3\langle 112\rangle$ superpartial (**Fig. A1(b)**). (c) In **Story 3**, Vorontsov *et al.* [7] (2012) suggested that two consecutive partial dislocations enter the γ′ phase on adjacent {111} planes, producing the two CSFs. Through thermally activated atomic reshuffling, the CSFs reorder to convert the APB into an SISF, resulting in an overall $a/3\langle 112\rangle$ superpartial (**Fig. A.1(c)**). (d) In **Story 4**, León-Cázares *et al.* [9] (2022) proposed that when two dissimilar channel dislocations meet at the γ/γ′ interface, then the cross-slip facilitates the formation of $a/3\langle 112\rangle$ superpartial to nucleate a nascent SISF (**Fig. A.1(d)**).

**Appendix 2. The corresponding interaction force and energy calculations**

Based on geometric model was established in **Fig. 12(a)**, the spatial relationship among these dislocations satisfies:

$$r_{2-3} = \sqrt{(r_{3-1} - cos\theta \cdot r_{2-1})^2 + (sin\theta \cdot r_{2-1})^2} \tag{A.1}$$

where $\theta = 70.53°$ is the angle between the $(\bar{1}11)$ plane and $(1\bar{1}1)$ planes, $r_{3-1} = 22\ nm$ is the distance between the leading and trailing partials within the ISF (measured from the HRSTEM image in **Fig. 6(c)**), $r_{2-1}$ is the distance between the leading Shockley partial ($\mathbf{30°, \beta C}$) and the perfect dislocation ($\mathbf{60°, CB}$), and $r_{2-3}$ is the distance between the trailing Shockley partial ($\mathbf{30°, D\beta}$) on the $(\bar{1}11)$ plane and the same perfect dislocation.

According to **Eqs. (15)** and **(17)**, the interaction forces $F_{2-1}$ and $F_{2-3}$ were calculated as functions of $r_{2-1}$. Along the vector direction $\boldsymbol{r_{2\_1}}$, the total interaction force is the sum of $F_{2-3}cos\alpha$ and $F_{2-1}$:

$$F_{total}^{ISF} = F_{2-3}cos\alpha + F_{2-1} \tag{A.2}$$

$$\alpha = 180° - \theta - \arcsin\left(\frac{sin\theta \cdot r_{2-1}}{r_{2-3}}\right) \tag{A.3}$$

In addition, based on **Eq. (16)**, we calculated the interaction energy between both leading ($\mathbf{30°, \beta C}$) and trailing ($\mathbf{30°, D\beta}$) partials and the perfect dislocation ($\mathbf{60°, CB}$). The total interaction energy ($E_{total}$) is written as:

$$E_{total}^{ISF} = E_{2-3} + E_{2-1} \tag{A.4}$$

In the ESF configuration, as illustrated in **Fig. 11(d)**, the relationship among the leading and trailing Shockley partials ($\mathbf{30°, D\beta}$ and $\mathbf{30°, 2D\beta}$) and the conjugate 60° mixed dislocation ($\mathbf{60°, BD}$) can be expressed as:

$$r_{5-6} = \sqrt{(r_{6-4} - cos\theta \cdot r_{5-4})^2 + (sin\theta \cdot r_{5-4})^2} \tag{A.5}$$

where $r_{6-4} = 21\ nm$ represents the distance between the leading and trailing partials within the ESF (based on ESF structure pinned γ/γ' interface from the HRSTEM image in **Fig. 4(b)**). $r_{5-4}$ is the separation between the leading Shockley partial (**30°, Dβ**) on the ($\bar{1}11$) plane and the perfect dislocation (**60°, BD**) on the conjugate ($1\bar{1}1$) plane, while $r_{5-6}$ denotes the distance between the two trailing Shockley partials (**30°, 2Dβ**) and the same perfect dislocation.

Due to the similarity of triangular geometric configuration with ISF and ESF, along the vector direction $\mathbf{r_{5\_4}}$, the total interaction force is the sum of $F_{5-6}cos\alpha$ and $F_{5-4}$, which can be expressed as:

$$F_{\text{total}} = F_{5-6}cos\alpha + F_{5-4} \tag{A.6}$$

$$\alpha = 180° - \theta - \arcsin\left(\frac{sin\theta \cdot r_{5-4}}{r_{5-6}}\right) \tag{A.7}$$

In addition, the total interaction energy ($E_{\text{total}}$) between both leading (**30°, Dβ**) and trailing (**30°, 2Dβ**) partials and the perfect dislocation (**60°, BD**) is written as:

$$E_{\text{total}} = E_{5-4} + E_{5-6} \tag{A.8}$$

**Appendix 3. Analytical model of Frank partial climb**

In this appendix, a theoretical framework is established to quantify the total driving force acting on a Frank partial and its corresponding climb velocity. The derivation is presented step-by-step: we first consider vacancy diffusion in an idealized homogeneous medium containing a single edge dislocation, and then specialize to the case of a Frank partial bounding a superlattice SF in the γ/γ' microstructure, as illustrated in **Fig. A.2**.

**A.3.1. An individual full edge dislocation: Coupling vacancy diffusion with dislocation climb**

We begin with a closed system including a single edge dislocation, which act as a sink/source for vacancies. This simple geometry allows us to couple a classical radial diffusion model for vacancies to the climb of the dislocation [39].

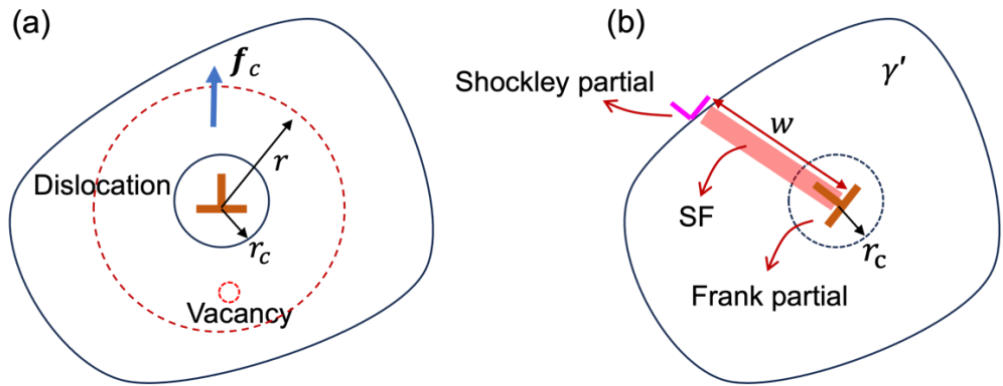

**Fig. A.2. Schematic illustration of the analytical model developed for Frank partial climb.** (a) Dislocation climbs under a radial vacancy flux pattern. $\boldsymbol{f}_c$ represents the climb component of the total driving force. (b) A Frank partial climbs in a superlattice SF configuration; an SESF case is shown here as an example.

Consider a cylindrical region surrounding the dislocation core, within which a radially symmetric diffusion field develops, as schematically illustrated in **Fig. A.2(a)**. In such a configuration, the steady-state radial vacancy flux can be expressed as:

$$j_r = \frac{A}{r} \tag{A.9}$$

where $r$ is the radial distance from the dislocation core, $A$ is a constant determined by boundary conditions.

Fick's first law for vacancy diffusion is:

$$j_r = -\frac{D_v \partial c_v(r)}{\partial r} \tag{A.10}$$

where $D_v$ is the vacancy diffusion coefficient, and $c_v(r)$ is the vacancy concentration at $r$. With boundary conditions are

$$c_v = c_v^0, \quad at \quad r = r_0 \tag{A.11.1}$$

$$c_v = c_v^c, \quad at \quad r = r_c \tag{A.11.2}$$

where $c_v^0$ is the reference vacancy concentration at the far-field $r_0$, $c_v^c$ is the vacancy concentration at the dislocation core region at the dislocation core radius $r_c$.

Substituting **Eq. (A.9)** into **Eq. (A.10)**, and applying boundary conditions in **Eq. (A.11)**, we derive the vacancy flux, $j_r$, as:

$$j_r = -\frac{D_v(c_v^c - c_v^0)}{r \ln(r_0/r_c)} \tag{A.12}$$

The vacancy concentration becomes:

$$c_v(r) = c_v^c + \frac{c_v^0 - c_v^c}{\ln(r_0/r_c)} \ln\left(\frac{r}{r_c}\right) \tag{A.13}$$

For an edge dislocation with a length of $l$ and a Burgers vector magnitude of $b$, climbing at velocity $v_c$, the total volume change associated with vacancy absorption (negative climb) or emission (positive climb) can be expressed as:

$$V^* = v_c b l \tag{A.14}$$

Meanwhile, mass conservation requires this volume change equals to the integrated vacancy flux across the cylindrical surface per unit time:

$$V^* = -j_r 2\pi r l \quad (A.15)$$

The negative sign is added to keep consistent with the fact that positive next flux into the core would lead to negative climb.

Combining **Eq. (A.14)** and **Eq. (A.15)**, we get:

$$v_c = -\frac{j_r 2\pi r}{b} = \frac{2\pi D_v}{b} \frac{c_v^0 - c_v^c}{\ln(r_0/r_c)} \quad (A.16)$$

**Eq. (A.16)** illustrates that, the climb velocity is determined by the net vacancy flux, i.e., the difference of vacancy concentrations between the near core region and the far field [39]. When $c_v^c > c_v^0$ (vacancy supersaturation at dislocation core), vacancies are emitted from the dislocation core into the lattice, the extra half-plane of the edge dislocation expands, dislocation climbs in the negative direction, i.e., $v_c < 0$. While, when $c_v^c < c_v^0$, vacancies are absorbed into the dislocation core, the extra half-plane of the edge dislocation retracts, and dislocation climbs upwards (positive climb), $v_c > 0$.

**A.3.2. Climb of a Frank partial dislocation in superlattice SF configuration**

Now we consider the climb of a Frank partial as observed in the present work. In a binary γ/γ′ microstructure, a superlattice SF configuration consists of a Shockley partial pinned at the γ/γ′ interface and a Frank partial that climbs into the γ′ phase, as schematically shown in **Fig. A.2(b).** These two partials bound a superlattice SF with energy density $\gamma_{SF}$. Depending on the Burgers vector of the Frank partial, the nature of SF corresponds either to an SISF or an SESF. Specifically, insertion of the extra half-plane associated with the Frank partial into the γ′ lattice produces an SESF; whereas, removal of the extra half-plane results in an SISF. In the following analysis, we take the SESF case as a representative example. Note that, the radially symmetric diffusion field introduced in the preceding subsection represents a first-order approximation. Although the γ/γ′ morphology and the interface-pinned Shockley partial disrupt perfect cylindrical symmetry, these geometric deviations do not alter the essential coupling between vacancy flux and Frank partial climb captured by the present model.

Gao and Cocks [39] have shown that three types of configurational force act on a dislocation that promote or inhibit its climb, including:
1) The climb component of the Peach-Koehler force, $f_{cl}$, due to the elastic field,

$$f_{cl} = \boldsymbol{f}_{PK} \cdot \boldsymbol{n} \quad (A.17)$$

where $f_{PK} = f_{app} + f_{self} + f_{inter}$ is the total Peach-Koehler force, arising from the externally applied stress, the dislocation self-stress and elastic interactions with other defects such as other dislocations and interfaces. The climb direction is $\boldsymbol{n} = \frac{\boldsymbol{b}_F \times \boldsymbol{l}_F}{|\boldsymbol{b}_F \times \boldsymbol{l}_F|}$, with $\boldsymbol{b}_F$ and $\boldsymbol{l}_F$ denoting the Burgers vector and the normalized line vector of the Frank partial, respectively.

The contribution from the externally applied stress is $(\boldsymbol{\sigma}_{app} \cdot \boldsymbol{b}_F) \times \boldsymbol{l}_F \cdot \boldsymbol{n}$. For the Frank partial considered here, $\boldsymbol{b}_F = a/3[1\bar{1}\bar{1}]$, $\boldsymbol{l}_F = [110]/\sqrt{2}$. Under a uniaxial compressive stress $\sigma_{33}$ applied in a polycrystalline specimen, the corresponding contribution to the climb force can be calculated as $\frac{1}{3}\sigma_{33}b_F$. This estimation provides a representative magnitude of the contribution from the externally applied stress; the actual stress experienced by a given Frank partial will vary with the corresponding local stress partitioning in the $\gamma'$ phase.

The contribution from the self-stress field of an edge Frank partial does not generate a net vacancy flux in the radial diffusion model adopted here. The hydrostatic component of the self-stress is antisymmetric with respect to the slip plane; the tensile and compressive components induce equal but opposite driving force on vacancies, and thus only resulting in a localized vacancy redistribution rather than a net vacancy flux.

The elastic interaction force per unit length between two parallel dislocations separated by a distance $w$ is given by [37]:

$$f_{inter}(w) = \frac{\mu_{\gamma'}}{2\pi w}(\boldsymbol{b}_S \cdot \boldsymbol{l}_S)(\boldsymbol{b}_F \cdot \boldsymbol{l}_F) + \frac{\mu_{\gamma'}}{2\pi w(1-\nu)}[(\boldsymbol{b}_S \times \boldsymbol{l}_S) \cdot (\boldsymbol{b}_F \times \boldsymbol{l}_F)] \quad (A.18)$$

where $\mu_{\gamma'}, \nu$ are the shear modulus and Poisson's ratio of the $\gamma'$ phase, respectively. For the SISF configuration, the Burgers vectors are $\boldsymbol{b}_S = \mathbf{D\beta} = a/6[211]$ (trailing Shockley partial) and $\boldsymbol{b}_F = \mathbf{\beta B} = a/3[1\bar{1}\bar{1}]$ (leading Frank partial). For the SESF configuration, the Burgers vectors are $\boldsymbol{b}_S = \mathbf{2D\beta} = a/3[211]$ (trailing Shockley partial) and $\boldsymbol{b}_F = \mathbf{B\beta} = a/3[1\bar{1}\bar{1}]$ (leading Frank partial). In both cases, the line directions of the two partials are identical, $\boldsymbol{l}_S = \boldsymbol{l}_F = [110]/\sqrt{2}$. Substituting these vectors into **Eq. (A.18)** shows that the scalar products entering the elastic term are zero, and thus the elastic interaction forces between the leading Frank and trailing Shockley partials are zero. Therefore, the climb component of total Peach-Koehler force can be simplified as,

$$f_{cl} = \frac{1}{3}\sigma_{33}b_F \quad (A.19)$$

2) The so-called osmotic force, $f_{os}$, arising from the change of free energy due to creation or annihilation of vacancies during climb. For an edge dislocation, the osmotic force per unit length is given as [39],

$$f_{os} = -\frac{kTb_F}{\Omega_f}\ln\left(\frac{c_v^c}{c_v^0}\right) \tag{A.20}$$

where $k$ is the Boltzman constant, $T$ is the absolute temperature. $\Omega_f$ is the vacancy formation volume.

3) The drag force, $f_{dr}$, opposes dislocation climb and is associated with the change in the free energy of the vacancies due to the evolving local stress field of a climbing dislocation. This term is often negligible compared to the previous two forces [39]. In the present configuration, however, an additional drag force arises from the energy cost of expanding or retracting the bounding superlattice SF as the Frank partial climbs, as the configuration shown in **Fig. A.2(b)**. We therefore treat the SF as an energy penalty on the expansion of the stacking fault. For an SESF, expansion of the fault corresponding to negative climb, and the drag force per unit length is

$$f_{dr} = \gamma_{SESF} \tag{A.21}$$

Note that, for an SISF, expansion of the SF corresponds to positive climb, so the sign is reversed, i.e., $f_{dr}(SISF) = -\gamma_{SISF}$.

The total configuration climb force per unit length on the Frank partial can then be summed as,

$$f_c = f_{cl} + f_{os} + f_{dr} \tag{A.22}$$

At an equilibrium state, where $f_c = 0$, we can derive the equilibrium vacancy concentration at dislocation core as,

$$c_v^{eq} = c_v^0 \exp\left(\frac{(f_{cl}+f_{dr})\Omega_f}{b_F kT}\right) \tag{A.23}$$

Vacancy diffusion along the dislocation core is typically very fast, so it is commonly assumed that the core concentration is maintained at its equilibrium value, $c_v^c = c_v^{eq}$, under all relevant conditions. While, outside the core, vacancy transport is controlled by much slower bulk diffusion. The resulting concentration gradient between the core and the matrix gives rise to a net flux of vacancies into or out of the core region, leading to dislocation climb until an equilibrium state is achieved.

By substituting **Eq. (A.23)** for $c_v^c$ into the general climb velocity expression, **Eq. (A.16)**, we derive the climb velocity as,

$$v_c = -\frac{j_r 2\pi r}{b} = \frac{2\pi D_v}{b \ln(r_0/r_c)} c_v^0 \left[1 - \exp\left(\frac{(f_{cl}+f_{dr})\Omega_f}{b_F kT}\right)\right] \tag{A.24}$$

This equation highlights two distinct contributions to the climb process: i) the perfector in front of the bracket, $\frac{2\pi D_v}{b \ln(r_0/r_c)}$, reflects the diffusivity of the vacancies. ii) The exponential term, $c_v^0 \left(1 - \exp\left(\frac{(f_{cl}+f_{dr})\Omega_f}{kT}\right)\right)$, captures the difference in vacancy chemical potential between the Frank partial core and far-filed, which determines the direction of the net vacancy flux, and thus the climb direction and rate.